\documentclass[prab,amsmath,amssymb,aps,twocolumn]{revtex4}

\usepackage{slashed}
\usepackage{hyperref}	
\usepackage{placeins}
\usepackage{amsthm}
\usepackage{adjustbox}
\usepackage{graphicx}
\usepackage{mathrsfs}	
\usepackage{soul}	
\usepackage{color}	
\usepackage{appendix}
\usepackage{eufrak}

\newcommand{\de}{\ensuremath{\partial}}
\newcommand{\dee}{\,\ensuremath{\textrm{d}}}

\newcommand{\inty}[4]{\ensuremath{ \int_{#1}^{#2} \! #3 \, \dee#4 }}

\newcommand{\field}[1]{\mathbb{#1}}
\newcommand{\ip}[2]{\ensuremath{ \left< \left. #1 \right| #2 \right> } }

\usepackage{enumitem, bm}
\usepackage[font=footnotesize,labelfont=bf]{caption}
\newcommand{\comment}[1]{}
\newcommand{\la}{\langle}
\newcommand{\ra}{\rangle}

\newcommand{\cJ}{\mathcal{J}}

\newcommand{\R}{\mathbb{R}}
\newcommand{\Z}{\mathbb{Z}}

\newcommand{\bra}[1]{\la #1|}
\newcommand{\ket}[1]{| #1 \ra}
\newcommand{\ketl}[1]{#1 \ra}

\DeclareMathOperator{\range}{range}

\DeclareMathOperator{\imag}{Im}

\usepackage{blkarray}

\allowdisplaybreaks

\renewcommand{\vec}[1]{\boldsymbol{#1}}
\graphicspath{{images/}}

\numberwithin{lemma}{section}
\numberwithin{example}{section}
\numberwithin{figure}{section}
\numberwithin{proposition}{section}
\numberwithin{equation}{section}
\numberwithin{theorem}{section}
\numberwithin{remark}{section}
\numberwithin{definition}{section}
\numberwithin{assumption}{section}

\usepackage{xcolor}
\definecolor{purp}{RGB}{160, 32, 240}
\definecolor{lightblue}{RGB}{32, 160, 240}

\usepackage{lipsum}

\begin{document}

\title{The Iterated Projected Position Algorithm for Constructing Exponentially Localized Generalized Wannier Functions for Periodic and Non-Periodic Insulators in Two Dimensions and Higher}

\author{Kevin D. Stubbs}
\affiliation{Department of Mathematics, Duke University, Box 90320, Durham, NC 27708, USA}
\email{kstubbs@math.duke.edu}
\author{Alexander B. Watson}
\affiliation{Department of Mathematics, University of Minnesota, Minneapolis, MN 55455, USA}
\email{watso860@umn.edu}
\author{Jianfeng Lu}
\affiliation{Department of Mathematics, Department of Physics, and Department of Chemistry, Duke University, Box 90320, Durham, NC 27708, USA}
\email{jianfeng@math.duke.edu}

\begin{abstract}
  Localized bases play an important role in understanding electronic structure. In periodic insulators, a natural choice of localized basis is given by the Wannier functions which depend a choice of unitary transform known as a gauge transformation. Over the past few decades, there have been many works which have focused on optimizing the choice of gauge so that the corresponding Wannier functions are maximally localized or reflect some symmetry of the underlying system. In this work, we consider fully non-periodic materials where the usual Wannier functions are not well defined and gauge optimization is impossible. To tackle the problem of calculating exponentially localized generalized Wannier functions in both periodic and non-periodic system we discuss the ``Iterated Projected Position (IPP)'' algorithm. The IPP algorithm is based on matrix diagonalization and therefore unlike optimization based approaches it does not require initialization and cannot get stuck at a local minimum. Furthermore, the IPP algorithm is guaranteed by a rigorous analysis to produce exponentially localized functions under certain mild assumptions. We numerically demonstrate that the IPP algorithm can be used to calculate exponentially localized bases for the Haldane model, the Kane-Mele model (in both $\Z_2$ invariant even and $\Z_2$ invariant odd phases), and the $p_x + i p_y$ model on a quasi-crystal lattice. 
  \end{abstract}
  

\maketitle

\section{Introduction}
When modeling electronic properties of materials, we often want to focus attention on a spectral subspace of an effective single-particle electronic Hamiltonian.
To do this, we must choose a basis (possibly other than the eigenfunctions themselves) to represent this subspace.
Not all bases are equally desirable however; bases which are well localized in space are particularly useful in both theoretical and computational studies \cite{1991Nenciu,1993King-SmithVanderbilt,2012MarzariMostofiYatesSouzaVanderbilt,2008MostofiYatesLeeSouzaVanderbiltMarzari,2020Wannier90}. 

In insulators (materials with a spectral gap at the Fermi level), the subspace of interest is the Fermi projection, the range of the Fermi projector $P$. When the insulator is periodic, a natural choice of localized basis is given by the Wannier functions, which are calculated by integrating a choice of Bloch basis with respect to the crystal quasi-momentum over the Brillouin zone. 
Wannier functions however depend on a choice of unitary transform on the Bloch functions known as a ``gauge transformation''. By making different choices of gauge, it is possible to change the localization properties of the corresponding Wannier functions. It is now known that under certain assumptions it is possible to pick the gauge on the Bloch functions so that the Wannier functions decay exponentially quickly away from their maximum value. These ``exponentially localized Wannier functions'' (ELWFs) play a central role in the study of periodic materials and the modern theory of polarization \cite{1993King-SmithVanderbilt,1994Resta,1999Goedecker,2005LeeNardelliMarzari,2006StengelSpaldin,2012MarzariMostofiYatesSouzaVanderbilt}. Because of the importance of ELWFs, much research over the past 30 years has been dedicated to understanding when it is possible to choose the gauge so that the corresponding Wannier functions are exponentially localized and how to compute such gauges.

For gapped periodic systems in one dimension there always exists a choice of gauge so that corresponding Wannier functions are exponentially localized \cite{1991Nenciu,1982NenciuNenciu}. In contrast, for gapped periodic systems in two and three dimensions ELWFs do not always exist. It is now understood that a choice of gauge corresponding to ELWFs exists if and only if certain topological invariants vanish \cite{2018MonacoPanatiPisanteTeufel}. In addition to these theoretical results, there has also been great progress with respect to numerical methods for calculating localized Wannier functions in periodic materials. In seminal work, Marzari and Vanderbilt proposed a numerical method based on gradient descent for optimizing the choice of gauge so that the resulting Wannier functions are as localized as possible \cite{1997MarzariVanderbilt}. Subsequently, this numerical method (and later refinements) were implemented into the software package \texttt{Wannier90} \cite{2020Wannier90}. One difficulty with the gradient descent procedure proposed by Marzari-Vanderbilt is that a poor choice of initialization can lead to Wannier functions which are not well localized. More recent work has looked at developing alternate optimization schemes to the one proposed by Marzari-Vanderbilt \cite{2015MustafaCohCohenLouie} or creating a good initial gauge choice by using symmetries in the underlying system \cite{2017CancesLevittPanatiStoltz}. There also have been propositions to generate localized Wannier function by using techniques from numerical linear algebra \cite{2015DamleLinYing,2017DamleLinYing}.


In this paper, we tackle the problem of constructing an exponentially localized basis for the Fermi projection for \textit{non-periodic} insulators 
and we refer to functions in any such basis as ``exponentially localized generalized Wannier functions'' (ELGWFs). We present the Iterated Projected Position (IPP) algorithm which we have proven in previous work constructs ELGWFs under fairly general assumptions \cite{2020StubbsWatsonLu}. The key difficulty in fully non-periodic systems is that Bloch functions do not exist. 
Therefore, we must find an alternate criterion (one which does not make reference to the Bloch functions
) 
for constructing ELGWFs. This problem has been solved in one dimension through the work of Kivelson \cite{1982Kivelson}, Niu \cite{1991Niu}, and Nenciu-Nenciu \cite{1998NenciuNenciu}. As the culmination of these works, it has been proven that in one dimension the eigenfunctions of the projected position operator $PXP$, where $P$ is the Fermi projector and $X$ is the position operator, are exponentially localized in both periodic and non-periodic systems. The IPP algorithm directly extends the work of Kivelson, Niu, and Nenciu-Nenciu to higher dimensions and is based on diagonalizing sequences of projected position operators. As a result of this, unlike methods which use optimization, such as Marzari-Vanderbilt functional minimization \cite{1997MarzariVanderbilt}, the IPP algorithm does not require any initial guesses and cannot get stuck at local minima. Like the eigenfunctions of $PXP$ in an infinite periodic system, the output ELGWFs of IPP are generally closed under lattice translations when the Hamiltonian is periodic, justifying the terminology ``generalized'' Wannier functions. 

We numerically demonstrate that the IPP algorithm can generate ELGWFs for systems with Dirichlet boundary conditions, periodic boundary conditions, time reversal symmetric systems (both $\Z_2$ invariant even and odd), and quasi-crystals. While we mainly focus on systems in two dimensions, the IPP algorithm easily generalizes to three dimensions (and higher) and provably produces ELGWFs under analogous assumptions to the two dimensional case.

The remainder of the paper is organized as follows. We begin by reviewing the definition of Wannier functions in periodic systems and the connection between ELWFs and projected position operators in Section \ref{sec:proj-pos-and-elwfs}. Having made this connection, we then introduce the iterated projected position (IPP) algorithm in Section \ref{sec:ipp-alg}. We give an overview of how IPP can be adapted to respect model symmetries in Section \ref{sec:other-ops} before giving details for periodic boundary conditions (Section \ref{sec:periodic}) and Bosonic and Fermionic time-reversal symmetries (Section \ref{sec:symmetry}). We explain how to intentionally break time-reversal symmetry as necessary in Section \ref{sec:tr_breaker}, and summarize these results in Section \ref{sec:results-summary}. 

After stating our main results, we turn to make connections between our results and previous work. In Section \ref{sec:mv-connection} we discuss the connection between the IPP algorithm and Marzari-Vanderbilt functional minimization and in Section \ref{sec:topo-connection} we discuss the connection between the IPP algorithm and the theory of topological invariants. Next, we test the IPP algorithm in a wide range of numerical tests in Section \ref{sec:numerics}. Finally, in Section \ref{sec:conclusions} we give an overview of our results and discuss future directions.

\section{Projected Position Operators and Exponentially Localized Wannier Functions}
\label{sec:proj-pos-and-elwfs}
As discussed previously, in periodic insulators a natural choice of localized basis is given by the Wannier functions. While the present work applies in \textbf{both} the periodic and non-periodic cases, it will be worthwhile to briefly review the basics of Wannier function theory in the periodic case to make connections with previous work more clear. 

For any periodic insulator with crystal lattice $\Lambda$, we can find an orthogonal basis of (generalized) eigenfunctions of the Hamiltonian which are also eigenfunctions of lattice translations. Such a basis of eigenfunctions is known as a Bloch basis and is denoted $\{ \psi_{n\vec{k}}(\vec{r}) \}$, where $n$ denotes the band index and $\vec{k}$ denotes the crystal quasi-momentum. In two dimensions, given a Bloch basis $\{ \psi_{n\vec{k}}(\vec{r}) \}$, for each $\vec{R} \in \Lambda$ the Wannier function centered at $\vec{R}$ is defined by the following integral over the Brillouin zone:
\begin{equation}
  \label{eq:wannier-function}
  w_{n\vec{R}}(\vec{r}) = \frac{1}{A} \int_{\text{BZ}} e^{-i \vec{k} \cdot \vec{R}} \psi_{n\vec{k}}(\vec{r}) \dee{\vec{k}}
\end{equation}
where $A$ is the area of the Brillouin zone.

Now recall that the eigenfunctions of $H$ are only defined up to a choice of complex phase. Hence, given a choice of $\{ \psi_{n\vec{k}}(\vec{r}) \}$, we could alternatively define the Wannier functions in Equation \eqref{eq:wannier-function} by making the substitution:
\[
    \psi_{n\vec{k}}(\vec{r}) \mapsto e^{i \lambda_{n\vec{k}}} \psi_{n\vec{k}}(\vec{r})
\]
where $\{ \lambda_{n\vec{k}} \}_{\vec{k} \in BZ} \subseteq \R$. More generally, for a system with $N$ bands, this degeneracy is defined by a collection of $N \times N$ unitary matrices $\{ U^{(\vec{k})} \}_{\vec{k} \in \text{BZ}}$ and substituting the following expression into Equation \eqref{eq:wannier-function}:
\begin{equation}
  \label{eq:gauge-freedom}
  \psi_{n\vec{k}}(\vec{r}) \mapsto \sum_{m} U_{nm}^{(\vec{k})} \psi_{m\vec{k}}(\vec{r}),
\end{equation}
which leaves the occupied subspace invariant. 

The mapping in Equation \eqref{eq:gauge-freedom} is known as a ``gauge transformation'' and an instance of the matrices $\{ U^{(\vec{k})} \}_{\vec{k} \in \text{BZ}}$ is known as ``choice of gauge''. By changing the choice of gauge, one can change whether the corresponding Wannier functions are localized in space or not. 

In \cite{1959Kohn}, Kohn proved that for inversion-symmetric crystals in one dimension with an isolated band there always exists a choice of gauge so that the corresponding Wannier functions decay \emph{exponentially fast} in space. This work was expanded on by Des Cloizeaux \cite{1964DesCloizeaux,1964DesCloizeaux2} and Nenciu-Nenciu \cite{1982NenciuNenciu} who proved that for arbitrary periodic insulators in one spatial dimension, there always exists a choice of gauge so that the Wannier functions are exponentially localized.
Having settled the question of existence of ELWFs for periodic systems in one dimension, it is natural to ask 
how the result generalizes to periodic insulators in higher dimensions. 
This question has been studied in detail by many authors and a full characterization of when a basis of ELWFs exists is now known in dimensions two and three. In two dimensions, ELWFs exist whenever the Chern number, a topological invariant associated to the Fermi projection, vanishes. In three dimensions, ELWFs exist whenever three ``Chern-like'' topological invariants associated to the Fermi projection all vanish
\cite{1964DesCloizeaux,1964DesCloizeaux2,1983Nenciu,1988HelfferSjostrand,1991Nenciu,2007BrouderPanatiCalandraMourougane,2007Panati,2018MonacoPanatiPisanteTeufel}. 

For systems where the lack of periodicity plays an important part in the material's properties (for example, in systems with defects or edges) far less is known. When a material is not periodic, Bloch theory does not apply so trying to find ELWFs by the usual methods of gauge optimization fails. Despite this, it has been conjectured that an exponentially localized basis for the Fermi projection should still exist \cite{1974KohnOnffroy,1991Niu,1993NenciuNenciu,2016CorneanHerbstNenciu} especially when the system in question is close to periodic. In fact, many of previous results about non-periodic Wannier functions were proved by perturbation or ``continuity''-type arguments \cite{1974KohnOnffroy,1993NenciuNenciu,1959Kohn,1991Nenciu,1993GellerKohn,1974RehrKohn, 2011WeinanLu}.

One approach to define Wannier functions in non-periodic materials was pioneered by Kivelson in \cite{1982Kivelson}. In this work, Kivelson proposed considering the eigenfunctions of the projected position operator, $PXP$, as non-periodic Wannier functions. To support this proposal, Kivelson showed that the exponentially localized Wannier functions found by Kohn in \cite{1959Kohn} are in fact eigenfunctions of $PXP$. Following up on the work by Kivelson, Niu argued in \cite{1991Niu} that in one dimension the eigenfunctions of $PXP$ should decay faster than any polynomial. A fully general, rigorous proof that the eigenfunctions of $PXP$ are exponentially localized in one dimension was finally given by Nenciu-Nenciu in \cite{1998NenciuNenciu}. The result by Nenciu-Nenciu is particularly powerful since it holds for an extremely wide class of systems, not just those which are close to being periodic. The IPP algorithm is an extension of the proposal of Kivelson, Niu, and Nenciu-Nenciu to higher dimensions.

\comment{In two dimensions, the main steps of IPP algorithm is to first diagonalize $P X P$ then use the results of this diagonalization to define a collection of orthogonal projectors $\{ P_j \}$. Once we define the projectors $\{ P_j \}$, the second step of the IPP algorithm is to diagonalize $P_j Y P_j$; hence the name the ``iterated projected position'' algorithm. An important feature of this method is that we can construct an ELGWFs \textit{without} needing to explicitly make a choice of gauge or run an optimization routine. We will give a more detailed account of the IPP algorithm in the next section.}

\section{Main Results}
\label{sec:results}

\subsection{The Iterated Projected Position Algorithm}
\label{sec:ipp-alg}
\begin{figure}
  \centering
  \includegraphics[width=.6\linewidth]{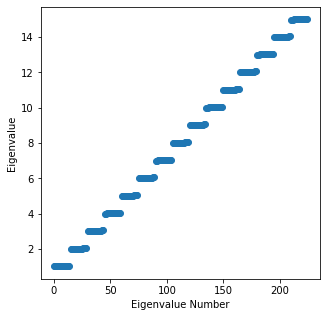}
  \caption{The sorted non-zero eigenvalues of the operator $PXP$ where $P$ is the Fermi projection for a non-topological Haldane model and $X$ is the standard position operator. The Haldane model was chosen with parameters $(v, t, t') = (3, 1,0.5)$ on a $12 \times 12$ system with Dirichlet boundary conditions (see Appendix \ref{sec:model-definitions} for definition of parameters).
  }
  \label{fig:hal_dir_evals_pxp}
\end{figure}

The key idea behind the IPP algorithm is the notion of \textit{uniform spectral gaps} (see Figure \ref{fig:hal_dir_evals_pxp} for a plot of the eigenvalues of an operator which has uniform spectral gaps). Informally speaking, an operator has uniform spectral gaps if its spectrum can be decomposed into a collection of disjoint sets $\{ \sigma_j \}_{j \in \cJ}$ which are separated by a minimum distance. The main result of our previous work \cite{2020StubbsWatsonLu} states that if $PXP$ has uniform spectral gaps then an exponentially localized basis for $\range{(P)}$ exists and the basis can be constructed by the IPP algorithm.

For a two dimensional system, this basis is constructed by the following steps. First, we let $X$ and $Y$ be a pair of position operators defined with respect to a pair of non-parallel coordinate axes. Next, we diagonalize the operator $PXP$ and assume that $PXP$ has uniform spectral gaps with decomposition $\{ \sigma_j \}_{j \in \cJ}$ \footnote{Note that if $PXP$ does not have uniform spectral gaps then the IPP algorithm fails.}. Given the decomposition $\{ \sigma_j \}$, by the spectral theorem, for each $\sigma_j$ we can construct an orthogonal projector, $P_j$, so that $P_j$ projects onto the span of the eigenvectors with eigenvalue from $\sigma_j$. Once we construct the projectors $\{ P_j \}_{j \in \cJ}$, the final step of the IPP algorithm is to diagonalize the operator $ P_j Y P_j$ for each $j \in \cJ$. It can be shown that the eigenfunctions of $P_j Y P_j$ are \textit{exponentially localized} in both $X$ and $Y$ simultaneously. To summarize these steps:
\begin{enumerate}[itemsep=-1ex]
\item Assume $PXP$ has uniform spectral gaps. 
\item Construct the projectors $\{ P_j \}_j$ for $PXP$.
\item For each $j$, diagonalize $P_j Y P_j$.
\end{enumerate}
In what follows, we will refer to applying steps 1-3 as ``applying the IPP algorithm using the sequence of position operators $X \rightarrow Y$''. We emphasize at this point that the spatial localization of the Wannier functions produced by the IPP algorithm relies purely on operator-theoretic estimates and hence does not require translation symmetry, in contrast to methods relying on Bloch function decomposition.

One can understand why the eigenfunctions of $P_j Y P_j$ are exponentially localized in both $X$ and $Y$ simultaneously by the following argument. Due to the separation between the different parts of the spectrum of $PXP$, using techniques from Combes-Thomas-Agmon theory \cite{1973CombesThomas}, it can be shown that the projectors $P_j$ are exponentially localized (i.e., as a matrix in spatial grid, the entries of $P_j$ decay exponentially quickly away from the diagonal). Since $P_j$ is also a spectral projector for $PXP$ it can also be shown that functions from $\range{(P_j)}$ are concentrated along a line of the form $x = \eta_j$ for some $\eta_j \in \R$. Since $P_j$ is concentrated along the line $x = \eta_j$, by restricting our focus to $\range{(P_j)}$ we have reduced the problem of finding ELGWFs in two dimensions to finding ELGWFs in ``essentially'' one dimension. But by reducing to a one dimensional problem, a generalization of the proof by Nenciu-Nenciu \cite{1998NenciuNenciu} shows that the eigenfunctions of $P_j Y P_j$ decay exponentially quickly in both $X$ and $Y$ simultaneously.

This argument easily generalizes to any dimension. For example in three dimensions, the sequence $X \rightarrow Y \rightarrow Z$ corresponds to the steps:
\begin{enumerate}[itemsep=-1ex]
\item Assume $PXP$ has uniform spectral gaps.
\item Construct the projectors $\{ P_{j_1} \}_{j_1}$ for $PXP$.
\item For each $j_1$, assume $P_{j_1} Y P_{j_1}$ has uniform spectral gaps.
\item Construct the projectors $\{ P_{j_1,j_2} \}_{j_2}$ for $P_{j_1} Y P_{j_1}$ for each $j_1$.
\item Diagonalize $P_{j_1,j_2} Z P_{j_1,j_2}$ for each $j_1,j_2$. 
\end{enumerate}
Furthermore, it can be rigorously proven that the eigenfunctions of $P_{j_1,j_2} Z P_{j_1,j_2}$ are exponentially localized in $X$, $Y$, and $Z$ simultaneously.


\subsection{Preserving and Breaking Symmetry in the IPP Algorithm}
\label{sec:other-ops}
Oftentimes we are not simply interested in constructing ELGWFs, we would also like to guarantee that these ELGWFs respect model symmetries such as periodic boundary conditions and time reversal symmetries.
The key for preserving or breaking such symmetries in the IPP algorithm lies in the choice of position operators.

Thus far, we have considered the sequence $X \rightarrow Y$ for a two dimensional system. So long as $PXP$ has uniform spectral gaps, the IPP algorithm will construct ELGWFs. However $X \rightarrow Y$ is not the only sequence of position operators which will result in the IPP algorithm constructing ELGWFs. For example, if we assume $PYP$ has uniform spectral gaps, then applying the IPP algorithm with the sequence $Y \rightarrow X$ will also construct ELGWFs. In fact, the proof from \cite{2020StubbsWatsonLu} generally implies that if $\tilde{X}$ and $\tilde{Y}$ are finite range, self-adjoint operators and $P \tilde{X} P$ has uniform spectral gaps, then applying the IPP algorithm with the sequence $\tilde{X} \rightarrow \tilde{Y}$ will construct a localized basis. By choosing $\tilde{X}$ and $\tilde{Y}$ to either respect or break certain symmetries we can force the results of the IPP algorithm to also preserve or break these symmetries. In this work, we demonstrate this principle by exhibiting sequences of position operators which lead to ELGWFs for a few specific combinations of boundary conditions and symmetries. 

With regards to boundary conditions, we will consider two kinds of boundary conditions: Dirichlet (open), where the electronic wave-function vanishes at the boundary of the computational domain, and periodic (closed). In the case of Dirichlet boundary conditions, there is no problem using $X \rightarrow Y$ as discussed in section \ref{sec:ipp-alg} to produce ELGWFs. We discuss operators which respect periodic boundary conditions in Section \ref{sec:periodic}. 

As for symmetries, although our primary focus is on methods which can be applied even when crystal lattice translation symmetries are broken, we will present operators such that the output of IPP respects this symmetry when it is present in section \ref{sec:periodic}. We will then discuss when the output of IPP respects two kinds of time-reversal symmetries: Bosonic and Fermionic, in section \ref{sec:symmetry}. We will finally discuss how to intentionally break Fermionic time reversal symmetry so that the output of IPP is exponentially localized even when there is a topological obstruction to existence of Wannier functions which are simultaneously exponentially localized and respectful of time-reversal symmetry in section \ref{sec:tr_breaker}.

\subsection{Periodic Position Operators}
\label{sec:periodic}

\subsubsection{Complex Exponential Position Operators} \label{sec:exp-per}

\begin{figure}
  \centering
  \includegraphics[width=\linewidth]{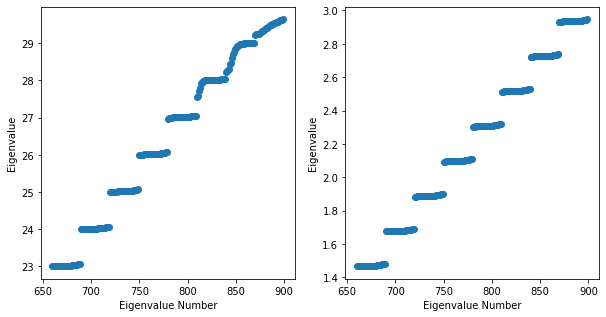}
  \caption{Plot of the largest $240$ non-zero eigenvalues of the operator $PXP$ (left) and $\imag{(\ln{(P e^{2 \pi i X / L_2} P)})}$ (right). Here $P$ denotes the Fermi projection for a non-topological Haldane model on a $30 \times 30$ system with periodic boundary conditions and $X$ is the standard position operator. The parameters used are $(v, t, t') = (3, 1,0.5)$  (see Appendix \ref{sec:model-definitions} for definition of parameters). Notice the last few gaps close in the left most plot but are uniformly spaced in the rightmost plot.}
  \label{fig:hal_per_evals_pxp}
\end{figure}

For finite systems with periodic boundary conditions, the standard position operators, $X$ and $Y$, are not the correct observables to measure position since these operators do not respect the boundary conditions. This fact is numerically present in the spectrum of the projected position operator $PXP$. In the left part of Figure \ref{fig:hal_per_evals_pxp}, we plot a subset of the sorted non-zero eigenvalues of the operator $PXP$ where $P$ is the Fermi projector for a non-topological Haldane model with periodic boundary conditions. In this Figure, we see that the last few gaps in the spectrum of $PXP$ close. While the IPP algorithm can still be applied in this case, the resulting ELGWFs will not be equally well localized (i.e. some of the generated functions will have significantly larger spread than the others). 

As suggested by Resta in \cite{1998Resta}, for a state $\ket{\psi}$ in a finite periodic material, its position in the $X$ direction is better defined using
\[
 \bar{x} = \frac{L_1}{2\pi} \imag{\ln{\bra{\psi} e^{2 \pi i X / L_1} \ket{\psi}}}
\]
where $L_1$ is the number of sites in the $X$ direction. This leads us to considering the sequence of projected position operators $P e^{2 \pi i X / L_1} P \rightarrow P e^{2 \pi i Y / L_2} P$ where at each step we sort the spectrum by taking the imaginary part of the natural logarithm of the eigenvalues. In the right part of Figure \ref{fig:hal_per_evals_pxp} we plot the spectrum of $\imag{(\ln{(P e^{2 \pi i X / L_2} P)})}$, where if $A$ is a diagonalizable matrix with $A = U D U^{-1}$ then $\imag(\ln{(A)}) := U \imag(\ln{(A)}) U^{-1}$. Notice that the spectrum shows clear uniform gaps.

One theoretical advantage of the complex exponential position operators is that whenever the system Hamiltonian has crystal lattice symmetry and $X$ and $Y$ are defined with respect to the crystal lattice basis vectors, the output of the IPP algorithm will respect this symmetry. To be more precise, let $\Lambda$ denote a two dimensional crystal lattice. If the Hamiltonian commutes with the translation operators
\begin{equation}
    T_{\vec{v}} f(\vec{r}) = f(\vec{r} + \vec{v})
\end{equation}
for every $\vec{v} \in \Lambda$, then the Wannier functions generated by the IPP algorithm will have the property that if $W(\vec{r})$ is a Wannier function, so are $T_{\vec{v}} W(\vec{r})$ for every $\vec{v} \in \Lambda$. 

To see this, first note that $H$ commuting with every $T_{\vec{v}}$ implies that $P$ does too. Now, if $\vec{a}_1$ and $\vec{a}_2$ are a basis of the crystal lattice and $X$ and $Y$ are defined with respect to $\vec{a}_1$ and $\vec{a}_2$, then $T_{\vec{a}_2}$ commutes with $P e^{2 \pi i X/L_1} P$ and $T_{\vec{a}_1} P e^{2 \pi i X/L_1} P = e^{2 \pi i/L_1} P e^{2 \pi i X/L_1} P T_{\vec{a}_1}$. It follows that $T_{\vec{a}_1} P_j = P_{j+1} T_{\vec{a}_1}$ and $T_{\vec{a}_2} P_j = P_j T_{\vec{a}_2}$. The conclusion now follows from $T_{\vec{a}_1} P_{j} e^{2 \pi i Y/L_2} P_j = P_{j+1} e^{2 \pi i Y/L_2} P_{j+1} T_{\vec{a}_1}$ and $T_{\vec{a}_2} P_j e^{2 \pi i Y/L_2} P_j = e^{2 \pi i/L_2} P_j e^{2 \pi i Y/L_2} P_j T_{\vec{a}_2}$. As a remark, note that the same argument holds for the standard position operators $X$ and $Y$ in an \emph{infinite} periodic system. 

One important difference between using the standard position operator $X$ and the complex exponential $e^{2 \pi i X / L_1}$ is that the projected position operator $P e^{2 \pi i X / L_1}P$ does not generally have orthogonal eigenvectors \footnote{Recall a matrix, $A$, has orthogonal eigenvectors if and only if it is normal. That is if $AA^\dagger = A^\dagger A$.}. While the projectors $P_j$ are still well defined in this case, they are not orthogonal projectors and numerically using these $P_j$ sometimes leads to trouble. To correct this issue, we apply L{\"o}wdin orthogonalization to the eigenvectors of $Pe^{2 \pi i X / L_1}P$ when we construct $P_j$ and when we construct the final results. This orthogonalization step has not been rigorously justified, but appears to work well numerically. We leave rigorously proving the correctness of using L{\"o}wdin orthogonalization to future work. This procedure is at least formally justified by the observation that for fixed values of $X$ and $Y$ we have
\begin{equation}
\begin{split}
    \frac{L_1}{2 \pi i} \left( e^{2 \pi i X / L_1} - 1 \right) \approx X, \quad \frac{L_2}{2 \pi i} \left( e^{2 \pi i Y / L_2} - 1 \right) \approx Y,
\end{split}
\end{equation}
as $L_j \rightarrow \infty$, $j = \{1,2\}$, and hence $P e^{2 \pi i X/L_1} P$ and $P e^{2 \pi i Y/L_2} P$ are approximately normal for large system sizes.

\subsubsection{Real Periodic Position Operators} \label{sec:cos-sin-ops}

An alternative to the sequence of complex exponential position operators which also respects periodic boundary conditions is the sequence
\begin{equation} \label{eq:cos-sin-ops}
    \begin{split}
    &\sin{(2 \pi X / L_1)} \rightarrow \cos{(2 \pi X / L_1)} \\ &\quad \rightarrow \sin{(2 \pi Y / L_2)} \rightarrow \cos{(2 \pi Y / L_2)}.
    \end{split}
\end{equation}
The intuition behind this sequence is the following. Recall that assuming $PXP$ has uniform spectral gaps, we can define the band projectors $\{ P_j \}$. Furthermore, for each $j$, functions from $\range{(P_j)}$ are concentrated along lines of the form $x = \eta_j$ for some $\eta_j \in \R$. Suppose that $P \sin{(2 \pi X / L_1)} P$ has uniform spectral gaps and let's denote the band projectors for $P \sin{(2 \pi X / L_1)} P$ as $\{ P_j^{\sin{}} \}$. Based on the previous analysis, We should expect that functions from $\range{(P_j^{\sin{}})}$ are concentrated along lines of the form $\sin{(2 \pi x / L_1)} = \eta_j$. Since $\sin{(2 \pi x / L_1)}$ is not injective for $x \in [0, L_1)$, generally the range of the projectors $\{ P_j^{\sin{}} \}$ will not be localized along a single line. To correct this issue, we note that the spectral projections of the operators $P_j^{\sin{}} \cos{(2 \pi X / L_1)} P_j^{\sin{}}$ are localized along a single line, and hence by including $\cos(2 \pi X/L_1)$ as well as $\sin(2 \pi X/L_1)$ in the sequence we obtain similar localization with respect to $x$ as with $PXP$. For the same reason, we must include both $\sin(2 \pi Y/L_2)$ and $\cos(2 \pi Y/L_2)$ in the sequence.

The sequence \eqref{eq:cos-sin-ops} has two advantages over the sequence of complex exponential position operators. First, since these operators are all self-adjoint, the theory from \cite{2020StubbsWatsonLu} \textit{does} directly apply and we can rigorously prove the functions produced by the IPP algorithm are exponentially localized. Second, unlike the complex exponentials, the operators \eqref{eq:cos-sin-ops} commute with time-reversal symmetry operators (see Section \ref{sec:symmetry}).

The sequence \eqref{eq:cos-sin-ops} has disadvantages relative to the sequence of complex exponentials. First, it is more complicated. Second, it does not have the property that when $H$ has crystal lattice symmetry, the Wannier functions produced by IPP retain that symmetry. Indeed, in practice we find that using complex exponentials gave better results in situations where preserving time-reversal symmetries is not important.



\subsection{Preserving Time Reversal Symmetries in the IPP algorithm}
\label{sec:symmetry}

In applications it may be important for Wannier functions to preserve time-reversal symmetries. In this work we consider two kinds of time-reversal symmetry.

The first time reversal symmetry we consider, which we refer to as Bosonic time-reversal symmetry, is the complex conjugation symmetry of models which neglect spin when the Hamiltonian is purely real. Specifically, define the anti-unitary complex conjugation operator $\mathcal{C}$ by
\begin{equation}
    \mathcal{C} f(\vec{r}) = \overline{ f(\vec{r}) }.
\end{equation}
Then we say Bosonic time-reversal symmetry holds whenever $\mathcal{C}$ commutes with the Hamiltonian $H$. In this case, we would like the Wannier functions to be invariant under $\mathcal{C}$, i.e. to be purely real.

The second time-reversal symmetry we will consider is Fermionic time-reversal symmetry. This is the symmetry of models which do account for spin, under the combined operation of complex conjugation and spin reversal. The anti-unitary operator $\Theta$ realizing this transformation satisfies, in contrast to $\mathcal{C}$, the condition
\begin{equation} \label{eq:FTR}
    \Theta^2 = - 1.
\end{equation}
In this case, we would like the the Wannier functions to be closed under $\Theta$ in the sense that if $W_1(\vec{r}),...,W_N(\vec{r})$ is the set of Wannier functions with centers closest to the origin (note $N$ must be even because of Kramers degeneracy) then there exists a unitary matrix $V$ (\eqref{eq:FTR} implies $V$ must also be skew-symmetric) such that
\begin{equation}
    \label{eq:fermi-trs}
    \begin{pmatrix} W_1(\vec{r}), ... , W_N(\vec{r}) \end{pmatrix}^\top = V \Theta \begin{pmatrix} W_1(\vec{r}), ... , W_N(\vec{r}) \end{pmatrix}^\top.
\end{equation}
It can happen that exponentially localized Wannier functions satisfying \eqref{eq:fermi-trs} do not exist. For periodic systems, it is known ELWFs satisfying \eqref{eq:fermi-trs} only exist when a $\mathbb{Z}_2$-valued topological invariant defined through the occupied Bloch functions (known as the $\mathbb{Z}_2$ invariant) vanishes \cite{2006FuKane,2017CorneanMonacoTeufel}. 

With appropriate choices of position operators, the IPP algorithm will automatically preserve the above time-reversal symmetries. We give short proofs in each case, starting with the case of Bosonic time-reversal symmetry.

Suppose $H$ commutes with $\mathcal{C}$, i.e., is purely real, and let $\tilde{X}$ and $\tilde{Y}$ denote \emph{real} position operators. Since the eigenvectors of a real Hermitian matrix can always be chosen to be real, we know that that the projector $P$ (which is a spectral projector for $H$) is also real. Since $\tilde{X}$ and $\tilde{Y}$ are real position operators, using this same reasoning we can conclude that $P \tilde{X} P$, $P_j$, and $P_j \tilde{Y} P_j$ are all real matrices. Therefore, the eigenfunctions of $P_j \tilde{Y} P_j$ can also be chosen to be real and hence Bosonic time reversal symmetry is preserved.

Now suppose $H$ commutes with $\Theta$, and let $\tilde{X}$ and $\tilde{Y}$ denote position operators which also commute with $\Theta$. It follows that $\Theta$ commutes with $P$, $P \tilde{X} P$, and $P_j \tilde{Y} P_j$. But now we have that $\Theta$ preserves the eigenspaces of $P_j \tilde{Y} P_j$, which is exactly \eqref{eq:fermi-trs}. 

We remark that it is easy to see that of the position operators already introduced, $X$, $Y$, and the real periodic position operators \eqref{eq:cos-sin-ops}, commute with both $\mathcal{C}$ and $\Theta$, while the complex exponential position operators do not.



\subsection{The Time Reversal Breaker $A_{TRB}$}
\label{sec:tr_breaker}


In the periodic case, it is well known that the existence of time reversal symmetry implies that there is a choice of Bloch gauge so that the Wannier functions are exponentially localized \cite{2007BrouderPanatiCalandraMourougane}. Unfortunately, as shown in \cite{2016FiorenzaMonacoPanati} when the $\Z_2$ invariant is non-zero, there cannot exist an orthogonal basis which is both exponentially localized and satisfies time reversal symmetry. Since the IPP algorithm preserves time reversal symmetry with the choice of position operators $X \rightarrow Y$ (see Section \ref{sec:symmetry}), the IPP algorithm using position operators $X$ and $Y$ must necessarily fail for $\Z_2$ invariant odd systems. 

To avoid this issue, inspired the work in by Silvestrelli, Marzari, Vanderbilt, and Parrinello~\cite{1998SilvestrelliMarzariVanderbiltParrinello}, we define a local, bounded, self-adjoint perturbation, $A_{TRB}$, which anti-commutes with time reversal symmetry and define the ``time reversal  broken'' position operators defined as follows:
\begin{equation}
    \begin{split}
    X_{TRB} := X + A_{TRB} \\
    Y_{TRB} := Y + A_{TRB}
    \end{split}
\end{equation}
Since $A_{TRB}$ anti-commutes with time reversal symmetry the position operators $X_{TRB}$ and $Y_{TRB}$ no longer commute with time reversal symmetry and hence the resulting eigenfunctions of the IPP algorithm will also break time reversal symmetry. Importantly, we can choose $A_{TRB}$ so that the theoretical results from \cite{2020StubbsWatsonLu} still imply that the output of the IPP algorithm is exponentially localized.

In our numerics, we test the Kane-Mele model which has four sites per unit cell ($A\uparrow$, $B\uparrow$, $A\downarrow$, $B\downarrow$) and therefore the position operator $X$ can be written as acting locally as follows (see Appendix \ref{sec:model-definitions} for more details on the Kane-Mele model):
\[
  (X\psi)_{m,n}
  =
  \begin{bmatrix}
    m\psi_{m,n}^{A\uparrow} \\[1ex]
    m\psi_{m,n}^{B\uparrow} \\[1ex]
    m\psi_{m,n}^{A\downarrow} \\[1ex]
    m\psi_{m,n}^{B\downarrow} 
  \end{bmatrix}
\]
We then introduce a term which couples the up and down spins at each site:
\[
  \begin{blockarray}{cccccc}
    && A\uparrow & B\uparrow & A\downarrow & B\downarrow \\[1ex]
    \begin{block}{cc[cccc]}
      A\uparrow &&  &  & 1 &  \\
      B\uparrow &&  &  &  & 1 \\
      A\downarrow && 1 &  &  &  \\
      B\downarrow &&  & 1  &  &  \\
    \end{block}
\end{blockarray}
\]
It's easy to check that this matrix is self-adjoint, has eigenvalues $\pm 1$, and that it anti-commutes with the time reversal operator. We propose adding these matrices to the original position operator $X$ at every site. That is,
\[
  ((X + A_{TRB})\psi)_{m,n}
  :=
  \left(
    m I_{4 \times 4}
  +
  \frac{1}{2}  
  \begin{bmatrix}
     &   & 1  &   \\
    &  &   & 1  \\
  1  &  &   &   \\
  & 1 &   &   
  \end{bmatrix}
\right)
\begin{bmatrix}
    \psi_{m,n}^{A\uparrow} \\[1ex]
    \psi_{m,n}^{B\uparrow} \\[1ex]
    \psi_{m,n}^{A\downarrow} \\[1ex]
    \psi_{m,n}^{B\downarrow} 
  \end{bmatrix}
\]
where $I_{4 \times 4}$ denotes a $4 \times 4$ identity matrix.

\begin{figure}
  \centering
 \includegraphics[width=\linewidth]{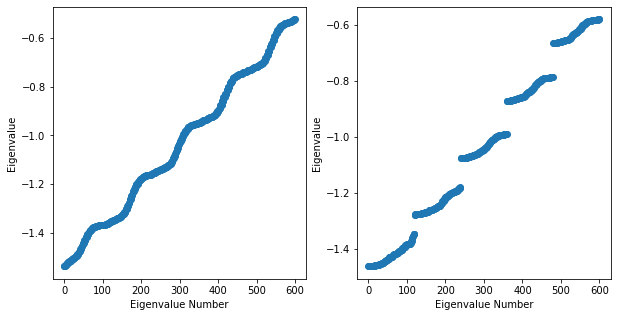}
  \caption{Plot of the first $600$ non-zero eigenvalues of $\arcsin{(P \sin{(2 \pi X / L_1)} P)}$ (left) and $\arcsin{(P \sin{(2 \pi X_{TRB} / L_1)} P)}$ (right). Here $P$ denotes the Fermi projection for a Kane-Mele model with odd $\Z_2$ invariant on a $30 \times 30$ system with periodic boundary conditions, $X$ is the standard position operator, and $X_{TRB} := X + A_{TRB}$. The parameters used are $(v, t, t', \lambda_R) = (4,1,0.6,0.5)$ (see Appendix \ref{sec:model-definitions} for definition of parameters). We observe that the addition of the time reversal breaker $A_{TRB}$ causes gaps in the spectrum to open.}
  \label{fig:km_per_evals_psinp_z2}
\end{figure}

As a note, the factor of $\frac{1}{2}$ ensures that the perturbation is small relative to the lattice spacing. In Figure \ref{fig:km_per_evals_psinp_z2} we compare the spectrum of $P \sin{(2 \pi X / L_1)} P $ and $P \sin{(2 \pi X_{TRB} / L_1)} P$ for a Kane-Mele model with odd $\Z_2$ invariant.



\subsection{Results Summary} 
\label{sec:results-summary}

In this section we have explained how to adapt the IPP algorithm so that the resulting Wannier functions have desired symmetry properties. In Section \ref{sec:numerics}, we will present numerical verifications that the methods of this section compute ELGWFs in the following cases.
\begin{enumerate}[itemsep=-1ex]
\item Dirichlet (open) boundary conditions, no time reversal symmetry.
\item Periodic boundary conditions, no time reversal symmetry.
\item Periodic boundary conditions, Bosonic time reversal symmetry holds.
\item Periodic boundary conditions, Fermionic time reversal symmetry holds, $\Z_2$ invariant even.
\item Periodic boundary conditions, Fermionic time reversal symmetry holds, $\Z_2$ invariant odd.
\end{enumerate}
In each case, we test the IPP algorithm with and without small random perturbations to the onsite potential. Note that such perturbations break translation symmetry and hence Wannier functions cannot be found using Bloch theory.
When we say that the $\Z_2$ invariant is even or odd, we refer to the $\Z_2$ invariant computed from the system without noise. In cases 1-4, the ELGWFs produced by the IPP algorithm respect boundary conditions and symmetries. In case 5, because of the presence of the $\field{Z}_2$ topological obstruction, to produce ELGWFs the IPP algorithm intentionally breaks Fermionic time-reversal symmetry.
We summarize the sequences of position operators used in each case in Table~\ref{table:summary}. 


\begin{table*}[t]
\begin{tabular}{lcp{.8\linewidth}}
\multicolumn{3}{l}{\textbf{\textit{No Time Reversal Symmetry}}} \\[1ex]
Dirichlet BCs & & $X \rightarrow Y$ \\[1ex]
  Periodic BCs & & $e^{2 \pi i X / L_1} \rightarrow e^{2 \pi i Y / L_2}$ \\[2.5ex]
  \multicolumn{3}{l}{\textbf{\textit{Time Reversal Symmetry, Periodic BCs}}} \\[1ex]
  Bosonic and $Z_2$ even  && $\sin{(2 \pi X / L_1)} \rightarrow \cos{(2 \pi X / L_1)} \rightarrow \sin{(2 \pi Y / L_2)} \rightarrow \cos{(2 \pi Y / L_2)}$ \\[1ex]
  $\Z_2$ odd & & $\sin{(2 \pi X_{TRB} / L_1)} \rightarrow \cos{(2 \pi X_{TRB} / L_1)} \rightarrow \sin{(2 \pi Y_{TRB} / L_2)} \rightarrow \cos{(2 \pi Y_{TRB} / L_2)}$
\end{tabular}
\caption{Summary of Main Results. Here $X_{TRB} := X + A_{TRB}$ and $Y_{TRB} := Y + A_{TRB}$ where $A_{TRB}$ is a local perturbation which breaks time-reversal symmetry. See Section \ref{sec:tr_breaker} for the definition of $A_{TRB}$ and discussion.
}
\label{table:summary}
\end{table*}

\section{Connection with Marzari-Vanderbilt Functional Minimization}
\label{sec:mv-connection}
Despite the fact that finding an exponentially localized basis is not always possible in two dimensions, in highly influential work Marzari and Vanderbilt \cite{1997MarzariVanderbilt} proposed choosing the gauge so that the variance of the resulting Wannier functions over the home unit cell is minimized. As noted in Marzari and Vanderbilt's original paper \cite{1997MarzariVanderbilt}, this variance functional can be separated into two parts: a part which depends on the choice of gauge and a part which is gauge invariant. Given a basis of Wannier functions on the home unit cell $\{ w_{n\vec{0}} \}$, a simple calculation shows that the gauge dependent part of the variance functional can be written as (see Appendix \ref{sec:mv-functional-calc} for more details):
\begin{equation}
  \label{eq:mv-functional-modified}
  \sum_{n} \| P (X - \mu_{n\vec{0}}^X) P  w_{n\vec{0}}\|^2 + \| P (Y - \mu_{n\vec{0}}^Y) P  w_{n\vec{0}}\|^2,
\end{equation}
where
\begin{equation}
    \mu_{n\vec{0}}^X := \bra{w_{n\vec{0}}} X \ket{w_{n\vec{0}}} \qquad \mu_{n\vec{0}}^Y := \bra{w_{n\vec{0}}} Y \ket{w_{n\vec{0}}}.
\end{equation}
Now notice that
\[
  \begin{split}
    \| P (X& - \mu_{n\vec{0}}^X) P  w_{n\vec{0}}\|^2 = 0 \\
    & \Longleftrightarrow P X P  w_{n\vec{0}} = \mu_{n\vec{0}}^X w_{n\vec{0}} \\
    & \Longleftrightarrow w_{n\vec{0}} \text{ is an eigenvector of $PXP$,}
  \end{split}
\]
where in the second line we have used that $w_{n\vec{0}} \in \range{(P)}$. This calculation shows that minimizing Equation \eqref{eq:mv-functional-modified} amounts to finding Wannier functions $\{ w_{n\vec{0}} \}_{n}$ which are approximately simultaneous eigenvectors of the operators $PXP$ and $PYP$. When $PXP$ and $PYP$ don't commute, finding a basis so that Equation \eqref{eq:mv-functional-modified} is exactly $0$ is impossible.

Unlike the Marzari-Vanderbilt approach, which tries to minimize both the $X$ and $Y$ directions simultaneously, the IPP algorithm takes a ``greedy'' approach to minimizing the objective in Equation \eqref{eq:mv-functional-modified} in the following sense. As noted previously by Kivelson \cite{1982Kivelson}, in one dimension the eigenvectors of $PXP$ can be thought of as the ``best'' approximation to an eigenstate of $X$ from $\range{(P)}$. From this perspective, we can interpret the sequence of diagonalizations $PXP \rightarrow P_j Y P_j$ as first trying to localize in $X$ among vectors in $\range{(P)}$ and then trying to localize in $Y$ among vectors in $\range{(P_j)}$. While there is no reason to expect that this iterative process will give a basis which is maximally localized, under the uniform spectral gaps assumption we can guarantee that this procedure gives a basis which is exponentially localized in both the $X$ and $Y$ directions simultaneously~\cite{2020StubbsWatsonLu}. 

\section{The Uniform Spectral Gaps Assumption implies Trivial Topology in Periodic Materials}
\label{sec:topo-connection}
In this section, we restrict to the special case of periodic systems so that we may make a direct connection between uniform spectral gaps and the theory of topological invariants. In particular, we will show that in two dimensions the uniform spectral gaps assumption implies that for general crystalline insulators the Chern number is zero and for crystalline insulators with Fermionic time reversal symmetry that the $\Z_2$-invariant is zero. Since in the crystalline case, it is now well understood that in two dimensions topological invariants completely characterize whether a basis of ELWFs exist or not \cite{2018MonacoPanatiPisanteTeufel}, the calculations in this section confirm that our results are consistent with existing theory.

The idea of connecting the spectrum of $PXP$ as used in the IPP algorithm to topological invariants is not new. This connection was first introduced by Soluyanov and Vanderbilt under the name of Wannier charge centers (WCCs) in the papers \cite{2011SoluyanovVanderbilt,2011SoluyanovVanderbilt2}. In Section \ref{sec:wcc-pxp} we will define the WCCs for a one dimensional system with a single band and connect the WCCs to the spectrum of $PXP$. While a one dimensional system with a single band is exceedingly simple, the techniques used in this simple example generalize easily to higher dimensions. In Section \ref{sec:usg-chern}, we will extend the construction of the WCCs to insulators with a single band in two dimensions and use the properties of the WCCs to show that uniform spectral gaps implies the Chern number must vanish. Finally, in Section \ref{sec:usg-z2}, using the WCCs we will show that uniform spectral gaps implies the $\Z_2$ invariant must vanish for an insulator with two bands and Fermionic time reversal symmetry.

While in the paper we only consider the simplest possible case (a single band for the Chern number and two bands for the $\Z_2$ invariant), the multi-band case follows by a similar argument by making some straightforward modifications. We direct interested readers to \cite[Appendix E]{2020StubbsWatsonLu} where the multi-band case is carefully worked out for the Chern number. Our presentation follows closely developments due in large part to Soluyanov and Vanderbilt \cite{2011SoluyanovVanderbilt2,2012Soluyanov,2017GreschYazyevTroyerVanderbiltBernevigSoluyanov}, who also show how to generalize the present ideas to more general cases. Our presentation also follows the mathematical works \cite{2016CorneanHerbstNenciu,2017CorneanMonacoTeufel}, where analytic and periodic Bloch function gauges are constructed rigorously.

\subsection{The Wannier Charge Centers and the Spectrum of $PXP$ in One Dimension}
\label{sec:wcc-pxp}
Let $L > 0$ be the one-dimensional lattice constant, and take the Brillouin zone as $k \in \left[-\frac{\pi}{L},\frac{\pi}{L}\right]$. We consider a single isolated Bloch band, denoting Bloch functions associated to the band by $\psi_{0k}$. One-dimensional Wannier functions are defined for each lattice vector $R = m L$ by
\begin{equation} \label{eq:1d_WF}
    W(r,R) = \frac{L}{2 \pi} \inty{-\frac{\pi}{L}}{\frac{\pi}{L}}{ \psi_{0k} e^{- i k R} }{k}.
\end{equation}
Niu \cite{1991Niu} (following \cite{1983Nenciu}, see also \cite{1997MarzariVanderbilt}) has explicitly displayed an analytic and periodic Bloch function gauge such that the associated one-dimensional Wannier functions are eigenfunctions of the operator $PXP$. We briefly review this construction now. We start by finding periodic Bloch functions $u(r,k) := e^{- i k r} \psi_{0k}$ in the adiabatic/parallel transport gauge \cite{Kato}, so that
\begin{equation} \label{eq:adiabatic}
    \ip{u(\cdot,k)}{\de_k u(\cdot,k)} = 0.
\end{equation}
This gauge makes the Bloch functions $\psi_{0k}$ analytic but not generally periodic in $k$. However, simplicity of the band implies that
\begin{equation}
    u\left(r,\frac{\pi}{L}\right) = e^{- i \frac{2 \pi}{L} x} \lambda u\left(r,-\frac{\pi}{L}\right)
\end{equation}
for some $\lambda \in \mathcal{U}(1)$, where $\mathcal{U}(N)$ denotes the group of $N \times N$ unitary matrices. To make the gauge periodic we replace $u(r,k)$ by
\begin{equation}
    e^{- i \frac{ \Gamma L }{ 2 \pi } k} u(r,k),
\end{equation}
where $\Gamma$ satisfies $e^{i \Gamma} = \lambda$. $\Gamma$ is not unique, since replacing $\Gamma$ by $\Gamma + 2 \pi m$ for any integer $m$ will also give an analytic and periodic gauge. However, shifting $\Gamma$ by $2 \pi$ is equivalent by shifting $R$ by one period in \eqref{eq:1d_WF}, so we may make the convention WLOG that $\Gamma \in [0,2\pi)$. Direct calculation using periodicity of the gauge and \eqref{eq:adiabatic} now shows that
\begin{equation} \label{eq:PXP_R}
    PXP W(r,R) = \left( R + \frac{\Gamma L}{2 \pi} \right) W(r,R).
\end{equation}
The quantity
\begin{equation}
    \overline{x} := \frac{\Gamma L}{2 \pi}
\end{equation}
is known as the Wannier charge center (WCC). Because of the non-uniqueness of $\Gamma$, $\overline{x}$ is defined only mod $2 \pi$. With our convention for $\Gamma$ however, we can assume WLOG that $\overline{x} \in [0,L)$. Since the choice of $R$ in \eqref{eq:1d_WF} was arbitrary, it follows that the spectrum of $PXP$ is $\sigma(PXP) = \field{Z} L + \overline{x}$, and hence $\overline{x}$ can be read easily from $\sigma(PXP)$. 



\subsection{Uniform Spectral Gaps implies Chern number is zero}
\label{sec:usg-chern}
In two dimensions we consider a crystal with lattice vectors $\vec{v}_1$, $\vec{v}_2$. We introduce spatial co-ordinates $(r_1,r_2)$ such that
\begin{equation}
    \vec{r} = \frac{ r_1 }{ L_1 } \vec{v}_1 + \frac{ r_2 }{ L_2 } \vec{v}_2,
\end{equation}
(here $L_j := |\vec{v}_j|, j = 1,2$) so that $(r_1,r_2) \in [0,L_1] \times [0,L_2]$ corresponds to a fundamental cell of the lattice $\Lambda$. Letting $\vec{w}_1$ and $\vec{w}_2$ denote dual vectors to $\vec{v}_1$ and $\vec{v}_2$ (such that $\vec{w}_i \cdot \vec{v}_j = 2 \pi \delta_{ij}$), we introduce $k$-space co-ordinates $(\kappa_1,\kappa_2)$ such that
\begin{equation}
    \vec{k} = \frac{ \kappa_1 }{2 \pi} \vec{w}_1 + \frac{ \kappa_2 }{2 \pi} \vec{w}_2,
\end{equation}
so that $\vec{k} := (\kappa_1,\kappa_2) \in [-\pi,\pi]^2$ corresponds to a fundamental cell (Brillouin zone) of the dual lattice $\Lambda^*$.

Assuming again a single isolated band, we can attempt to construct an analytic and periodic Bloch function gauge over the whole Brillouin zone in 2d by iterating the 1d construction detailed above. We start by constructing an analytic and periodic (with respect to $\kappa_2$) gauge along the line $(0,\kappa_2)$ where $\kappa_2 \in [-\pi,\pi]$ by exactly mimicking the 1d construction. We now extend this gauge to the whole Brillouin zone by parallel transporting the periodic Bloch functions $u(\vec{r},0,\kappa_2)$ along the lines $(\kappa_1,\kappa_2)$ where $\kappa_1 \in [-\pi,\pi]$ for each $\kappa_2$. In this way we construct a Bloch function gauge over the whole Brillouin zone which is analytic with respect to $\kappa_1$ and $\kappa_2$, but periodic only with respect to $\kappa_2$. Using simplicity of the band, we have that
\begin{equation}
    u(\vec{r}_1,\pi,\kappa_2) = e^{- i \vec{w}_1 \cdot \vec{r}} \lambda(\kappa_2) u(\vec{r},-\pi,\kappa_2),
\end{equation}
where $\lambda(\kappa_2) \in \mathcal{U}(1)$, and $\lambda(\kappa_2 + 2 \pi) = \lambda(\kappa_2)$ by periodicity of the gauge with respect to $\kappa_2$. 

We can try to ``mend'' the gauge by replacing the Bloch functions $u(\vec{r},\kappa_1,\kappa_2)$ along each line of constant $\kappa_2$ by
\begin{equation} \label{eq:trans}
    e^{- i \frac{ \Gamma(\kappa_2) }{ 2 \pi } \kappa_1} u(\vec{r},\kappa_1,\kappa_2),
\end{equation}
where $e^{i \Gamma(\kappa_2)} = \lambda(\kappa_2)$ for each $\kappa_2$ and the map $\kappa_2 \mapsto \Gamma(\kappa_2)$ is assumed analytic. The result of this process is a new ``mended'' gauge which is analytic with respect to $\kappa_1$ and $\kappa_2$ and periodic with respect to $\kappa_1$. The gauge will retain periodicity with respect to $\kappa_2$ if $\Gamma(\kappa_2) = \Gamma(-\kappa_2)$. It is possible that this does not hold despite the periodicity of $\lambda(\kappa_2)$, since this only implies $\Gamma(\pi) = \Gamma(-\pi)$ mod $2 \pi$. By identifying the ends of the Brillouin zone it is natural to view the map $\kappa_2 \mapsto e^{i \Gamma(\kappa_2)}$ as mapping $S^1 \rightarrow S^1$. From this perspective, the mended gauge will retain periodicity with respect to $\kappa_2$ if and only if the winding number of this map is zero.

\begin{figure}
  \centering
  \includegraphics[width=\linewidth]{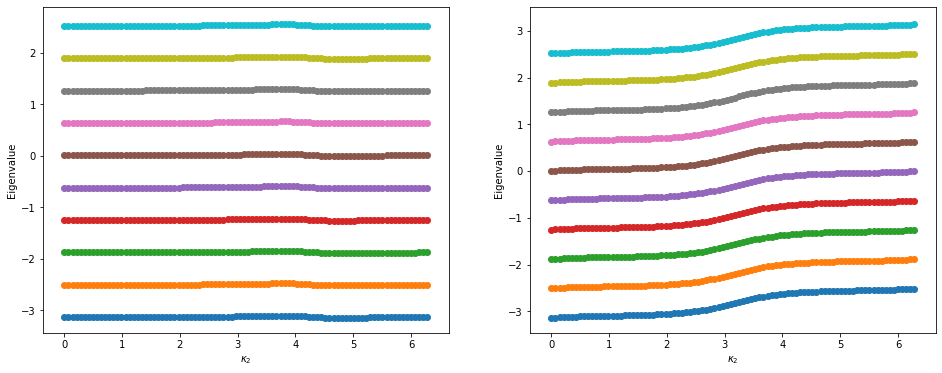}
  \caption{Wannier charge centers $\overline{x}_m(k_2)$ of the Haldane model with size $10 \times 90$ with parameters such that the Chern number is zero $(v, t, t') = (3,1,0.5)$ (left) and non-zero $(v, t, t') = (0,1,0.5)$ (right).
  }
  \label{fig:hal_per_bands_psinp}
\end{figure}

Non-trivial winding of the map $\kappa_2 \mapsto e^{i \Gamma(\kappa_2)}$ can be detected from the spectrum of the operator $P X P$ as follows. We define hybrid Wannier functions (HWFs) for each $R_1 = m L_1$ where $m \in \field{Z}$ by
\begin{equation}
    H(\vec{r},R_1,\kappa_2) = \frac{1}{2 \pi} \inty{-\pi}{\pi}{ \psi_{\kappa_1,\kappa_2}(\vec{r}) e^{- i \kappa_1 R_1} }{\kappa_1}.
\end{equation}
Then, letting $P(\kappa_2)$ denote the projection onto the Bloch functions along the line $(\kappa_1,\kappa_2)$ where $\kappa_1 \in [-\pi,\pi]$ for each $\kappa_2$ and adopting the gauge just constructed we have by essentially the same calculation leading to \eqref{eq:PXP_R},
\begin{equation}
\begin{split}
    &P(\kappa_2) X P(\kappa_2) H(r_1,r_2,R_1,\kappa_2) \\
    &= \left( R_1 + \overline{x}(\kappa_2) \right) H(r_1,r_2,R_1,\kappa_2), \\
\end{split}
\end{equation}
where $\overline{x}(\kappa_2) := \frac{ \Gamma(\kappa_2) L_1 }{ 2 \pi }$ can be understood as the WCC ``at $\kappa_2$''. Since $R_1$ is arbitrary, we see that the spectrum of $P(\kappa_2) X P(\kappa_2)$ is $\sigma(P(\kappa_2) X P(\kappa_2)) = \field{Z} L_1 + \overline{x}(\kappa_2)$. 
It is clear that if the map $\kappa_2 \mapsto e^{i \Gamma(\kappa_2)}$ winds, the map $\kappa_2 \mapsto \overline{x}(\kappa_2)$ must sweep out the whole interval $[0,L_1]$, and hence the spectrum of $PXP$, given by
\begin{equation}
    \sigma(PXP) = \bigcup_{\kappa_2 \in [-\pi,\pi]} \sigma( P(\kappa_2) X P(\kappa_2) ),
\end{equation}
cannot have spectral gaps. It follows that the uniform spectral gap assumption on $PXP$ implies that the mended Bloch function gauge constructed above is actually analytic and periodic in $\kappa_1$ and $\kappa_2$, from which ELWFs can be constructed via the usual construction. In Figure \ref{fig:hal_per_bands_psinp} we plot the imaginary part of the natural logarithm of $\sigma(P(\kappa_2) e^{2 \pi i X / L} P(\kappa_2))$ as $\kappa_2$ is varied showing different possible behaviors\footnote{As discussed in Section \ref{sec:periodic}, $P(\kappa_2) X P(\kappa_2)$ shows boundary effects in finite systems stemming from the fact that $X$ does not commute with lattice translations.}


We finally link these observations to the Chern number. Noting that in the gauge constructed above the Berry connection is 
\begin{equation}
    i \ip{u(\cdot,\kappa_1,\kappa_2)}{\de_{\kappa_1} u(\cdot,\kappa_1,\kappa_2) } = \frac{\Gamma(\kappa_2)}{2 \pi},
\end{equation}
we have, using Stokes' theorem (recall that the mended gauge is always analytic in $\kappa_1$ and $\kappa_2$ and periodic with respect to $\kappa_1$), that the Chern number 
\begin{equation}
    \mathcal{C} = \frac{1}{2 \pi} \left[ \Gamma(\pi) - \Gamma(-\pi) \right].
\end{equation}
Hence whenever $PXP$ has spectral gaps the Chern number must vanish.

\subsection{Uniform Spectral Gaps implies $\Z_2$ invariant is zero}
\label{sec:usg-z2}
We now consider the same two-dimensional setup as the previous section with the additional assumption that Fermionic time-reversal symmetry holds, i.e. that there exists an anti-unitary operator $\Theta$ such that $\Theta^2 = - 1$ and $\Theta P(\kappa_1,\kappa_2) \Theta = P(-\kappa_1,-\kappa_2)$ where $P(\kappa_1,\kappa_2)$ denotes the projection onto the set of Bloch functions at $(\kappa_1,\kappa_2)$. We assume the simplest possible case in this setting, which is of two Bloch bands isolated from the other Bloch bands of the Hamiltonian, and attempt to construct a Bloch function gauge $\vec{k} \mapsto \left( u(\vec{r},\vec{k}), v(\vec{r},\vec{k}) \right)$ which is analytic, periodic, and respects time-reversal symmetry in the sense that
\begin{equation} \label{eq:TR_sym}
    \begin{pmatrix} u(\vec{r},-\vec{k}) \\ v(\vec{r},-\vec{k}) \end{pmatrix} = V \Theta \begin{pmatrix} u(\vec{r},\vec{k}) \\ v(\vec{r},\vec{k}) \end{pmatrix}, \quad V := \begin{pmatrix} 0 & -1 \\ 1 & 0 \end{pmatrix}
\end{equation}
for all $\vec{k}$ in the Brillouin zone.

Given an arbitrary periodic Bloch function at $(0,0)$ $u(\vec{r},0,0)$, we let $v(\vec{r},0,0) := \Theta u(\vec{r},0,0)$. We can generate analytic and periodic Bloch functions $u(\vec{r},0,\kappa_2)$ and $v(\vec{r},0,\kappa_2)$ along the line $(0,\kappa_2), \kappa_2 \in [-\pi,\pi]$ via the 1d parallel transport procedure as in the previous sections. Analysis of the unitary realizing parallel transport shows that this gauge also respects \eqref{eq:TR_sym}. By extending this gauge to the whole Brillouin zone via parallel transport along the lines $(\kappa_1,\kappa_2), \kappa_1 \in [-\pi,\pi]$ for each fixed $\kappa_2$, we obtain a gauge which is analytic in $\kappa_1$ and $\kappa_2$, periodic with respect to $\kappa_2$, and respectful of time-reversal symmetry \eqref{eq:TR_sym}. It follows that
\begin{equation}
    \begin{pmatrix} u(\vec{r},\pi,\kappa_2) \\ v(\vec{r},\pi,\kappa_2) \end{pmatrix} = e^{- i \vec{w}_1 \cdot \vec{r}} U(\kappa_2) \begin{pmatrix} u(\vec{r},-\pi,\kappa_2) \\ v(\vec{r},-\pi,\kappa_2) \end{pmatrix},
\end{equation}
where $U(\kappa_2) \in \mathcal{U}(2)$ is analytic in $\kappa_2$, periodic in the sense that $U(\kappa_2 + 2 \pi) = U(\kappa_2)$ for every $\kappa_2$, and satisfies the symmetry
\begin{equation} \label{eq:this}
    U(\kappa_2) = ( V \Theta )^{-1} U^\dagger(-\kappa_2) V \Theta.
\end{equation}
By rotating the set of periodic Bloch functions along the line $(0,\kappa_2), \kappa_2 \in [-\pi,\pi]$, we can assume that $U(\kappa_2)$ is diagonal, with analytic and periodic eigenvalues $\lambda_u(\kappa_2), \lambda_v(\kappa_2) \in \mathcal{U}(1)$ such that $\lambda_v(-\kappa_2) = \lambda_u(\kappa_2)$, $\lambda_u(-\kappa_2) = \lambda_v(\kappa_2)$ for all $\kappa_2$ (by \eqref{eq:this}). In particular, we have $\lambda_u(0) = \lambda_v(0)$. Combining \eqref{eq:this} with periodicity of $U(\kappa_2)$ implies that $\lambda_u(\pm \pi) = \lambda_v(\pm \pi)$.

Just as in the case without time-reversal symmetry, we can attempt to ``mend'' the gauge so it is periodic with respect to $\kappa_1$ by replacing $u(\vec{r},\kappa_1,\kappa_2)$ by $e^{- i \frac{ \Gamma_u(\kappa_2) }{ 2 \pi } \kappa_1} u(\vec{r},\kappa_1,\kappa_2)$, where $\Gamma_u(\kappa_2)$ is chosen analytically in $\kappa_2$ such that $e^{i \Gamma_u(\kappa_2)} = \lambda_u(\kappa_2)$ for each $\kappa_2$ (and the same for $v(\vec{r},\kappa_1,\kappa_2)$). For the mended gauge to retain time-reversal symmetry we must have $\Gamma_v(-\kappa_2) = \Gamma_u(\kappa_2)$ and $\Gamma_u(-\kappa_2) = \Gamma_v(\kappa_2)$ (and hence $\Gamma_u(0) = \Gamma_v(0)$), while the degeneracies of $\lambda_u(\kappa_2)$ and $\lambda_v(\kappa_2)$ at $0$ and $\pm \pi$ ensure that $\Gamma_u(\pi) = \Gamma_v(\pi)$ and $\Gamma_u(-\pi) = \Gamma_v(-\pi)$ mod $2 \pi$. For the gauge to retain periodicity in $\kappa_2$, we require the additional conditions 
\begin{equation}
    \Gamma_u(\pi) = \Gamma_u(-\pi) \text{ and } \Gamma_v(\pi) = \Gamma_v(-\pi).
\end{equation}
Assuming we have chosen the gauge to respect time-reversal symmetry, these conditions are equivalent to
\begin{equation}
    \Gamma_u(\pi) = \Gamma_v(\pi) \text{ and } \Gamma_u(-\pi) = \Gamma_v(-\pi),
\end{equation}
although the second condition is clearly redundant. Recall that $\lambda_u(\pm \pi) = \lambda_v(\pm \pi)$ and hence $\Gamma_u(\pm \pi) = \Gamma_v(\pm \pi)$ mod $2 \pi$. Just as in the case without time-reversal symmetry, we can consider the maps $\kappa_2 \mapsto e^{i \Gamma_u(\kappa_2)}$, $\kappa_2 \mapsto e^{i \Gamma_v(\kappa_2)}$ as mapping $S^1 \rightarrow S^1$, and conclude that the mending process yields an analytic and periodic gauge which respects time-reversal symmetry if and only if the winding numbers of these maps are both zero (clearly they are equal up to a sign).

\begin{figure}
  \centering
  \includegraphics[width=\linewidth]{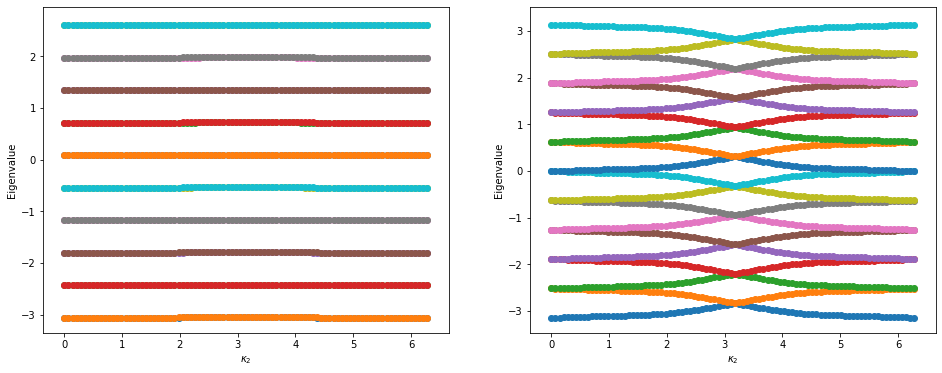}
  \caption{Wannier charge centers $\overline{x}_m(k_2)$ of the Kane-Mele model with size $10 \times 90$ with parameters such that the $\field{Z}_2$ index is zero $(v, t, t', \lambda_R) = (4,1,0.6,0.5)$ (left) and non-zero $(v, t, t', \lambda_R) = (0,1,0.6,0.5)$ (right).}
  \label{fig:km_per_bands_psinp}
\end{figure}

\begin{figure}
  \centering
  \includegraphics[width=.6\linewidth]{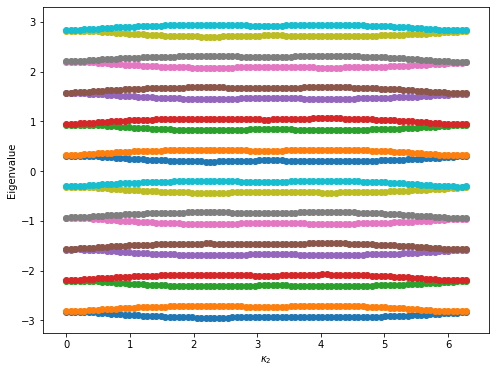}
  \caption{Wannier charge centers $\overline{x}_m(k_2)$ with the inclusion of $A_{TRB}$ of the Kane-Mele model with size $10 \times 90$ with parameters $(v, t, t', \lambda_R) = (0,1,0.6,0.5)$ ($\Z_2$ invariant odd)}
  \label{fig:km_per_bands_psinp_trb}
\end{figure}

We can again link the failure of the mending process to the spectrum of the operator $PXP$ as follows. Define HWFs for each $R_1 = m L_1$ where $m \in \field{Z}$ and each $\omega \in \{ u,v \}$ by
\begin{equation}
    H_\omega(\vec{r},R_1,\kappa_2) = \frac{1}{2 \pi} \inty{-\pi}{\pi}{ \psi_{0\vec{k},\omega}(\vec{r}) e^{- i \kappa_1 R_1} }{\kappa_1},
\end{equation}
where $\psi_{0\vec{k},\omega}(\vec{r}) = e^{i \vec{k} \cdot \vec{r}} \omega(\vec{r},\vec{k})$, and where each periodic Bloch function is assumed to be in the mended gauge defined above. By essentially the same calculation as in \eqref{eq:PXP_R} we have
\begin{equation}
\begin{split}
    &P(\kappa_2) X P(\kappa_2) H_\omega(\vec{r},R_1,\kappa_2)   \\
    &= ( R_1 + \overline{x}_\omega(\kappa_2) ) H_\omega(\vec{r},R_1,\kappa_2),
\end{split}
\end{equation}
where $\overline{x}_\omega(\kappa_2) := \frac{ \Gamma_\omega(\kappa_2) L_1 }{ 2 \pi }$. Since $R_1$ is arbitrary, we have $\sigma( P(\kappa_2) X P(\kappa_2) ) = \left[ \field{Z} L_1 + \overline{x}_u(\kappa_2) \right] \bigcup \left[ \field{Z} L_1 + \overline{x}_v(\kappa_2) \right]$. It is clear that if the maps $\kappa_2 \mapsto e^{i \Gamma_\omega(\kappa_2)}$ wind, the maps $\kappa_2 \mapsto \overline{x}_\omega(\kappa_2)$ must sweep out the whole interval $[0,L_1]$, and hence the spectrum of $PXP$ cannot have spectral gaps. It follows that the spectral gap assumption we make on $PXP$ implies the existence of an analytic, periodic, and time-reversal symmetric gauge over the whole Brillouin zone, and hence time-reversal symmetry-respecting ELWFs by the usual construction. Plots of $\sigma(P(\kappa_2) X P(\kappa_2))$ as $\kappa_2$ is varied showing different possible behaviors when time-reversal symmetry holds are shown in Figure \ref{fig:km_per_bands_psinp}. These figures should be compared with the same figures when $X$ is replaced by $X + A_{TRB}$ where $A_{TRB}$ does not respect time-reversal symmetry Figure \ref{fig:km_per_bands_psinp_trb}.

We finally link these observations to the $\field{Z}_2$ invariant. Noting that in the gauge constructed above the Berry connection takes the form
\begin{equation}
    \ip{ \omega(\cdot,\kappa_1,\kappa_2) }{ \de_{\kappa_1} \omega(\cdot,\kappa_1,\kappa_2) } = \frac{ \Gamma_\omega(\kappa_2) }{2 \pi} \quad \omega \in \{u,v\},
\end{equation}
Fu and Kane's definition of the $\field{Z}_2$ invariant in terms of time-reversal polarization $\Delta$ \cite{2006FuKane} becomes 
\begin{equation}
    \Delta = \frac{1}{2 \pi} \left( \left[\Gamma_u(\pi) - \Gamma_v(\pi)\right] - \left[\Gamma_u(0) - \Gamma_v(0)\right] \right) \text{ mod } 2.
\end{equation}
Since we have already fixed a gauge where $\Gamma_u(0) = \Gamma_v(0)$ and established that whenever $PXP$ has gaps we have $\Gamma_u(\pi) = \Gamma_v(\pi)$ we see that $\Delta$ vanishes.

\FloatBarrier

\section{Numerical Results}
\label{sec:numerics}
We now turn to numerically test our method for the Haldane, Kane-Mele, and $p_x + i p_y$ models (which we carefully define in Appendix \ref{sec:model-definitions}). We will test these models in the following scenarios:
\begin{enumerate}
\item Dirichlet Boundary Conditions (Section \ref{sec:dirichlet-bcs})
\begin{enumerate}[topsep=0ex]
\item Kane-Mele model with Dirichlet boundary conditions (Section \ref{sec:km-dir})
\item Kane-Mele model with Dirichlet boundary conditions with weak disorder (Section \ref{sec:km-dir-wk-dis})
\item $p_x + i p_y$ model with Dirichlet boundary conditions (Section \ref{sec:px-ipy-dir})
\end{enumerate}
\item Periodic Boundary Conditions (Section \ref{sec:periodic-bcs})
\begin{enumerate}[topsep=0ex]
\item Haldane Model with Periodic boundary conditions (Section \ref{sec:hm-per})
\item Haldane Model with Periodic boundary conditions with weak disorder (Section \ref{sec:hm-per-wk-dis})
\item Haldane Model with Periodic boundary conditions with strong disorder (Section \ref{sec:hm-per-str-dis})
\end{enumerate}
\item Time Reversal Symmetries (Section \ref{sec:tr-numerics})
\begin{enumerate}[topsep=0ex]
\item Haldane Model with Periodic boundary conditions and Bosonic time reversal symmetry (Section \ref{sec:hm-per-bos})
\item Kane-Mele Model with Periodic boundary conditions and $\Z_2$ invariant even (Section \ref{sec:km-per-z0})
\item Kane-Mele Model with Periodic boundary conditions and $\Z_2$ invariant odd (Section \ref{sec:km-per-z1})
\item Kane-Mele Model with Periodic boundary conditions, $\Z_2$ invariant even, and weak noise (Section \ref{sec:km-per-z1-wk-dis})
\end{enumerate}
\end{enumerate}
To demonstrate the effectiveness of our algorithm, we will display a number of plots which show the exponential decay of the generated orthonormal basis.

For the Haldane and Kane-Mele models we will run our tests on a $30 \times 30$ system and make plots of the following matrix. Here $\| \psi_{n,m} \|_2$ denotes the Euclidean norm of the sites in the $(n,m)$ cell:
\[
  \begin{bmatrix}
    \|\psi_{1,1}\|_2 & \|\psi_{1,2}\|_2 & \cdots & \|\psi_{1,30}\| \\[2ex]
    \|\psi_{2,1}\|_2 & \|\psi_{2,2}\|_2 & \cdots & \|\psi_{2,30}\| \\[2ex]
    \vdots & \vdots & \ddots & \vdots \\[2ex]
    \|\psi_{30,1}\|_2 & \|\psi_{30,2}\|_2 & \cdots & \|\psi_{30,30}\|_2 \\[2ex]
  \end{bmatrix}.
\]
We will plot this matrix as both a 3D surface plot as well as 2D intensity plot on a log scale.

Since the Ammann-Beekner tiling is not a lattice, we cannot easily translate our results for the $p_x + i p_y$ model to a matrix as we can for the Haldane and Kane-Mele models. For this model, we will instead plot the points in Ammann-Beekner quasi-lattice and at each point superimpose a circle whose radius is proportional to the Euclidean norm of the generalized Wannier function at that site (for an example of this, see Figure \ref{fig:px_ipy_ex}).

To verify the robustness of our algorithm, in some of our experiments we will randomly perturb the original Hamiltonian, $H$, by ``on-site disorder''. More specifically, we will consider the disordered Hamiltonian, $H_{disorder}$, as follows (where $\{ \ket{i} \}$ denotes the position basis):
\begin{equation}
  \label{eq:disorder}
  \begin{split}
  & H_{disordered} = H + \sum_{i} \eta_i \ket{i}\bra{i}, \\
  & \eta_i \overset{i.i.d}{\sim} \mathcal{N}(0, \sigma^2).\
  \end{split}
\end{equation}
That is, the disorder adds independent draws from a Gaussian distribution with mean $0$ and variance $\sigma^2$ to the diagonal entries of the original Hamiltonian.

\subsection{A Comment on Diagonalizing Projected Position Operators}
While theoretically it is convenient to work with projected position operators of the form $\tilde{P} \tilde{X} \tilde{P}$, for numerical purposes this matrix is quite large and computing all of the eigenvectors and eigenvalues of $\tilde{P} \tilde{X} \tilde{P}$ is wasteful when the projector $\tilde{P}$ is low rank. At every step in our numerics, we have access to a matrix with orthonormal columns, $\tilde{B}$, so that $\tilde{P} = \tilde{B} \tilde{B}^\dagger$. A simple calculation shows that if $v$ is an eigenfunction of $\tilde{B}^\dagger \tilde{X} \tilde{B}$ then $\tilde{B} v$ is an eigenfunction of $\tilde{P} \tilde{X} \tilde{P}$. Since the matrix $\tilde{B}^\dagger \tilde{X} \tilde{B}$ is significantly smaller than $\tilde{P} \tilde{X} \tilde{P}$, in our tests of the IPP algorithm we diagonalize the small matrix $\tilde{B}^\dagger \tilde{X} \tilde{B}$ to find the non-trivial eigenfunctions of $\tilde{P} \tilde{X} \tilde{P}$.

\subsection{Dirichlet Boundary Conditions}
\label{sec:dirichlet-bcs}

\subsubsection{Kane-Mele Model, Dirichlet Boundary Conditions}
\label{sec:km-dir}
\begin{figure}[h]
  \includegraphics[width=.7\linewidth]{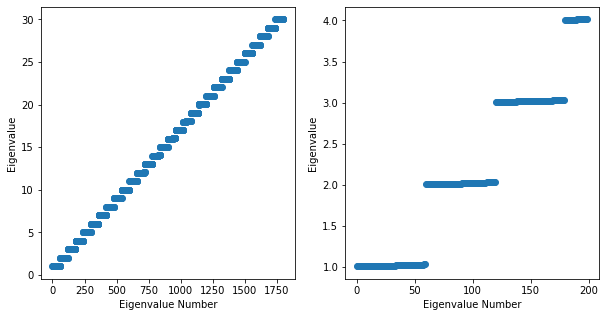}
  \caption{Plot of all non-zero eigenvalues (left) and the first $200$ non-zero eigenvalues (right) of $PXP$. Here $P$ denotes the Fermi projection for a Kane-Mele model on a $30 \times 30$ system with Dirichlet boundary conditions and $X$ is the standard position operator. The parameters used are $(v, t, t', \lambda_R) = (4,1,0.6,0.5)$  (see Appendix \ref{sec:model-definitions} for definition of parameters).}
  \label{fig:km_dir_evals_pxp}
\end{figure}

\begin{figure}[h]
  \includegraphics[width=.75\linewidth]{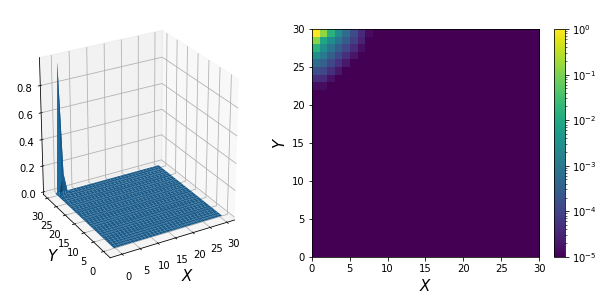} \\
  \includegraphics[width=.75\linewidth]{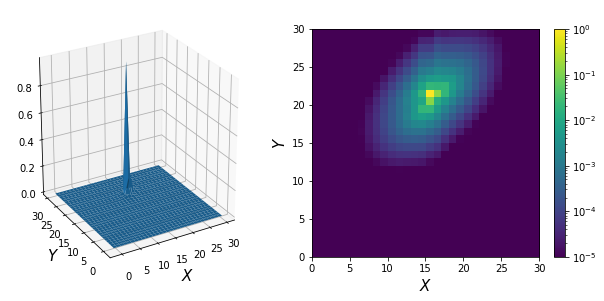}
  \caption{Plot of two of the generalized Wannier functions generated by the IPP algorithm using $X \rightarrow Y$ for the system from Figure \ref{fig:km_dir_evals_pxp}. A 3D surface plot of the results (left) and the corresponding 2D log plot (right).}
  \label{fig:km_dir_ex}
\end{figure}

As a first numerical example, let us consider the Kane-Mele model with Dirichlet boundary condition and parameters $(v,t,t',\lambda_R) = (4, 1, 0.6, 0.5)$. In Figure \ref{fig:km_dir_evals_pxp}, we plot all of the non-zero eigenvalues of $PXP$ (left) and the first $200$ non-zero eigenvalues of $PXP$ (right); notice the eigenvalues of $PXP$ has clear spectral gaps. In Figure \ref{fig:km_dir_ex}, we plot some of the eigenfunctions of $P_j Y P_j$; notice they are clearly exponentially localized.

\FloatBarrier

\subsubsection{Kane-Mele Model, Dirichet Boundary Conditions, Weak Disorder}
\label{sec:km-dir-wk-dis}
\begin{figure}[h]
  \includegraphics[width=.7\linewidth]{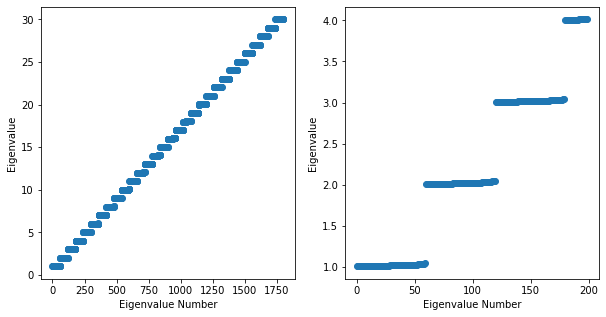}
  \caption{Plot of all non-zero eigenvalues (left) and the first $200$ non-zero eigenvalues (right) of $PXP$. Here $P$ denotes the Fermi projection for a Kane-Mele model with on-site disorder on a $30 \times 30$ system with Dirichlet boundary conditions and $X$ is the standard position operator. The parameters used are $(v, t, t', \lambda_R) = (4,1,0.6,0.5)$ and the disorder variance is $\sigma^2 = 0.5$ (see Appendix \ref{sec:model-definitions} for definition of parameters and Equation \eqref{eq:disorder} for definition of on-site disorder).}
  \label{fig:km_dir_wk_dis_evals_pxp}
\end{figure}

\begin{figure}[h]
  \includegraphics[width=.75\linewidth]{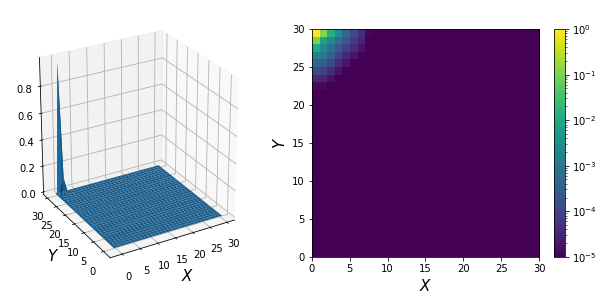} \\
  \includegraphics[width=.75\linewidth]{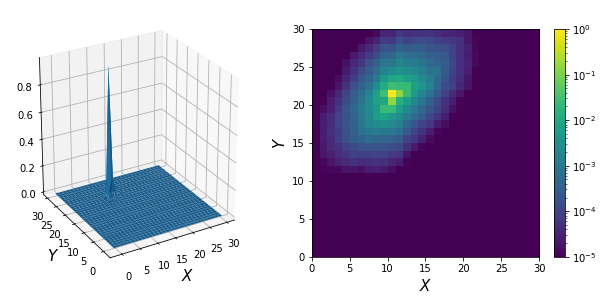}
  \caption{Plot of two of the generalized Wannier functions generated by the IPP algorithm using $X \rightarrow Y$ for the system from Figure \ref{fig:km_dir_wk_dis_evals_pxp}. A 3D surface plot of the results (left) and the corresponding 2D log plot (right).}
  \label{fig:km_dir_wk_dis_ex}
\end{figure}

Next, let's consider the same system as in Section \ref{sec:km-dir} with the addition of on-site disorder. The parameters for this model are $(v,t,t',\lambda_R) = (4, 1, 0.6, 0.5)$ and the on-site disorder has variance $\sigma^2 = 0.5$. In Figure \ref{fig:km_dir_wk_dis_evals_pxp}, we plot all of the non-zero eigenvalues of $PXP$ (left) and the first $200$ non-zero eigenvalues of $PXP$ (right); notice that the spectral gaps in $PXP$ are still present with weak disorder. In Figure \ref{fig:km_dir_wk_dis_ex}, we plot some of the eigenfunctions of $P_j Y P_j$.

\FloatBarrier
\subsubsection{$p_x + i p_y$ model, Dirichlet Boundary Conditions}
\label{sec:px-ipy-dir}
\begin{figure}[h]
  \includegraphics[width=.7\linewidth]{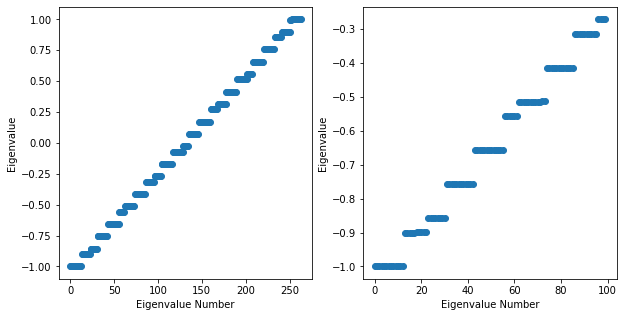}
  \caption{Plot of all non-zero eigenvalues (left) and the first $100$ non-zero eigenvalues (right) of $PXP$. Here $P$ denotes the Fermi projection for $p_x + i p_y$ model on the Ammann-Beekner lattice and $X$ is the standard position operator. The parameters used are $(\mu, t, \Delta) = (3,0.5, 1)$ (see Appendix \ref{sec:model-definitions} for definition of parameters).}
  \label{fig:px_ipy_evals_pxp}
\end{figure}

\begin{figure}[h]
  \includegraphics[width=.45\linewidth]{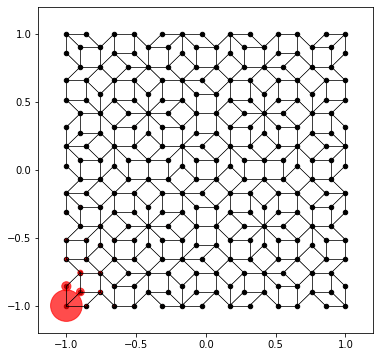} 
  \includegraphics[width=.45\linewidth]{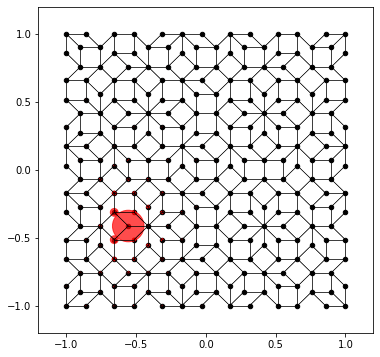} 
  \caption{Plot of two of the generalized Wannier functions generated by the IPP algorithm using $X \rightarrow Y$ for the system from Figure \ref{fig:px_ipy_evals_pxp}. The radius of the circle at each point in the lattice is proportional to the Euclidean norm of the generalized Wannier function at that site.}
  \label{fig:px_ipy_ex}
\end{figure}

Since the IPP algorithm does not make any assumptions about the underlying symmetries in the system, we can easily apply it to quasi-lattice systems such as the $p_x + i p_y$ model on the Ammann-Beekner lattice. In the following tests we choose parameters $(\mu, t, \Delta) = (3,0.5, 1)$ so that the gap in the Hamiltonian opens and the system is non-topological. In Figure \ref{fig:px_ipy_evals_pxp}, we see that $PXP$ has clear gaps and in Figure \ref{fig:px_ipy_ex} we see that the eigenfunctions of $P_j Y P_j$ are exponentially localized about their center.

\FloatBarrier

\subsection{Periodic Boundary Conditions}
\label{sec:periodic-bcs}
Now that we've tested Dirichlet boundary conditions, we move on to test periodic systems. For these tests, we will use the sequence of position operators $e^{2 \pi i X / L_1} \rightarrow e^{2 \pi i Y / L_1}$.

\subsubsection{Haldane Model, Periodic Boundary Conditions}
\label{sec:hm-per}
\begin{figure}[h]
  \includegraphics[width=.7\linewidth]{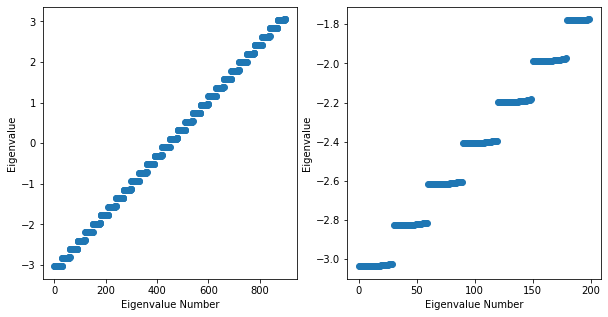} 
  \caption{Plot of all non-zero eigenvalues (left) and the first $200$ non-zero eigenvalues (right) of $\imag{\log{(P e^{2 \pi i X / L_1} P)}}$. Here $P$ denotes the Fermi projection for a Haldane model on a $30 \times 30$ system with periodic boundary conditions and $X$ is the standard position operator. The parameters used are $(v, t, t') = (3, 1,0.5)$ (see Appendix \ref{sec:model-definitions} for definition of parameters).}
  \label{fig:hm_per_evals_pexpp}
\end{figure}

We first consider a non-topological Haldane model with periodic boundary conditions and parameters $(v,t,t') = (3, 1, 0.5)$. In Figure \ref{fig:hm_per_evals_pexpp}, we plot all of the non-zero eigenvalues of $\imag{\log{(P e^{2 \pi i X / L_1} P)}}$ (left) and the first $200$ non-zero eigenvalues of $\imag{\log{(P e^{2 \pi i X / L_1} P)}}$(right); notice these eigenvalues have clear spectral gaps.

\begin{figure}[h]
  \includegraphics[width=.75\linewidth]{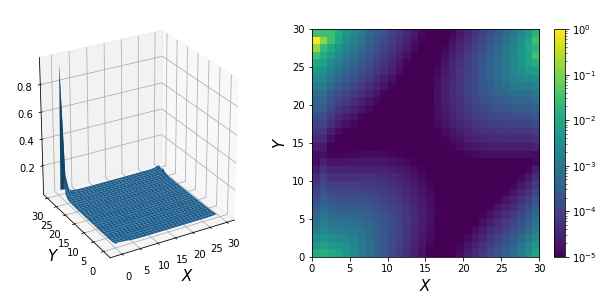} \\
  \includegraphics[width=.75\linewidth]{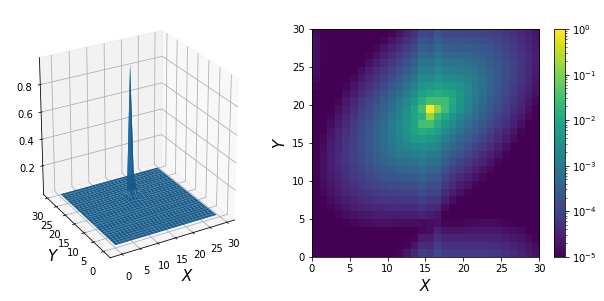} 
  \caption{Plot of two of the generalized Wannier functions generated by the IPP algorithm using $e^{2 \pi i X / L_1} \rightarrow e^{2 \pi i Y / L_1}$ for the system from Figure \ref{fig:hm_per_evals_pexpp}. A 3D surface plot of the results (left) and the corresponding 2D log plot (right).}
  \label{fig:hm_per_ex}
\end{figure}

\FloatBarrier
\subsubsection{Haldane Model, Periodic Boundary Conditions, Weak Disorder}
\label{sec:hm-per-wk-dis}
When we add any amount of unstructured disorder to a periodic system, the periodicity in the system is lost and therefore Bloch theory does not apply. Despite this issue, the IPP algorithm is robust to disorder so still it produces ELGWFs.

\begin{figure}[h]
  \includegraphics[width=.7\linewidth]{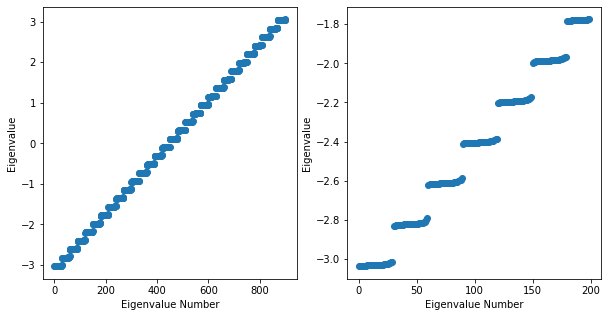}
  \caption{Plot of all non-zero eigenvalues (left) and the first $200$ non-zero eigenvalues (right) of $\imag{\log{(P e^{2 \pi i X / L_1} P)}}$. Here $P$ denotes the Fermi projection for a Haldane model with on-site disorder on a $30 \times 30$ system with periodic boundary conditions and $X$ is the standard position operator. The parameters used are $(v, t, t') = (3, 1,0.5)$ and the disorder variance is $\sigma^2 = 0.5$ (see Appendix \ref{sec:model-definitions} for definition of parameters and Equation \eqref{eq:disorder} for definition of on-site disorder).}
  \label{fig:hm_per_wk_dis_evals_pexpp}
\end{figure}

\begin{figure}[h]
  \includegraphics[width=.75\linewidth]{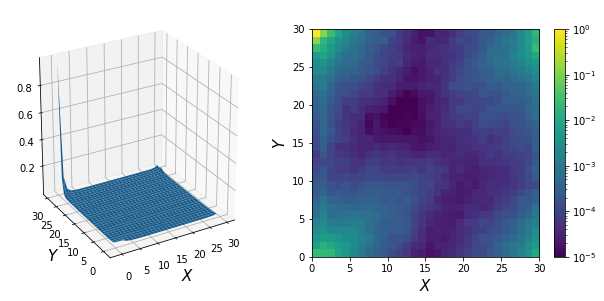} \\
  \includegraphics[width=.75\linewidth]{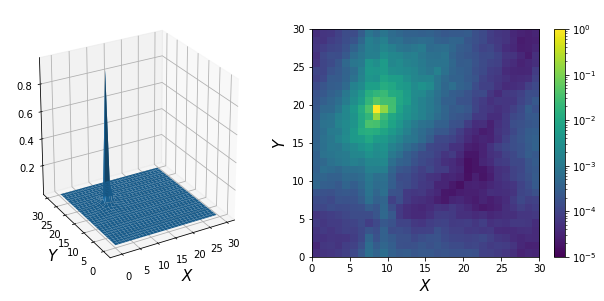}
  \caption{Plot of two of the generalized Wannier functions generated by the IPP algorithm using $e^{2 \pi i X / L_1} \rightarrow e^{2 \pi i Y / L_1}$ for the system from Figure \ref{fig:hm_per_wk_dis_evals_pexpp}. A 3D surface plot of the results (left) and the corresponding 2D log plot (right).}
  \label{fig:hm_per_wk_dis_ex}
\end{figure}

To numerically show this, let's consider the Haldane model with parameters $(v,t,t') = (3, 1, 0.5)$ as in Section \ref{sec:hm-per} with the addition of on-site disorder with variance $\sigma^2 = .5$. In Figure \ref{fig:hm_per_wk_dis_evals_pexpp}, we plot all of the non-zero eigenvalues (left) and the first $200$ non-zero eigenvalues of $\imag{\log{(P e^{2 \pi i X / L_1} P)}}$ (right); notice that the spectral gaps are still present with weak disorder. In Figure \ref{fig:hm_per_wk_dis_ex}, we plot a few of the results of the IPP algorithm.

\FloatBarrier
\subsubsection{Haldane Model, Periodic Boundary Conditions, Strong Disorder}
\label{sec:hm-per-str-dis}
\begin{figure}[h]
  \includegraphics[width=.7\linewidth]{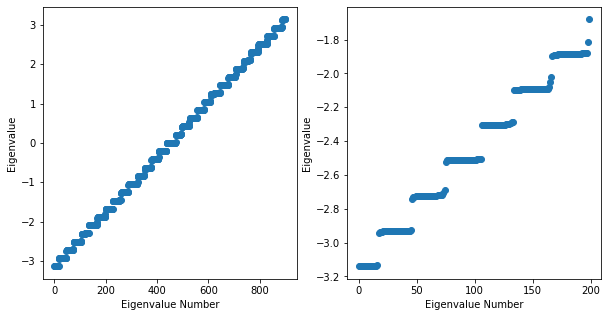} 
  \caption{Plot of all non-zero eigenvalues (left) and the first $200$ non-zero eigenvalues (right) of $\imag{\log{(P e^{2 \pi i X / L_1} P)}}$. Here $P$ denotes the Fermi projection for a Haldane model with on-site disorder on a $30 \times 30$ system with periodic boundary conditions and $X$ is the standard position operator. The parameters used are $(v, t, t') = (3, 1,0.5)$ and the disorder variance is $\sigma^2 = 100$ (see Appendix \ref{sec:model-definitions} for definition of parameters and Equation \eqref{eq:disorder} for definition of on-site disorder).}
  \label{fig:hal_per_str_dis_evals_pexpp}
\end{figure}

\begin{figure}[h]
  \includegraphics[width=.75\linewidth]{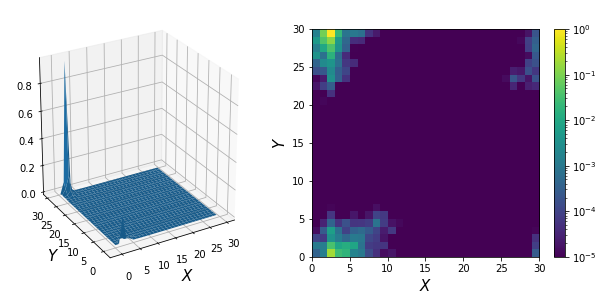} \\
  \includegraphics[width=.75\linewidth]{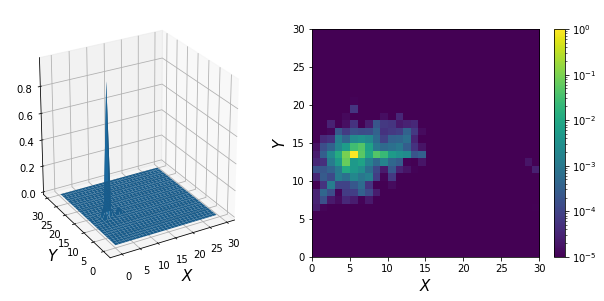}
  \caption{Plot of two of the generalized Wannier functions generated by the IPP algorithm using $e^{2 \pi i X / L_1} \rightarrow e^{2 \pi i Y / L_1}$ for the system from Figure \ref{fig:hal_per_str_dis_evals_pexpp}. A 3D surface plot of the results (left) and the corresponding 2D log plot (right).}
  \label{fig:hal_per_str_dis_ex}
\end{figure}

Next, we consider a periodic Haldane model with extremely strong on-site disorder. Although the gap of the Hamiltonian closes in this case, due Anderson localization \cite{1958Anderson}, we should expect that there still exists an exponentially localized basis for the Fermi projection. The parameters for this model are $(v,t,t') = (3, 1, 0.5)$ and the on-site disorder has variance $\sigma^2 = 100$. In Figure \ref{fig:hal_per_str_dis_evals_pexpp}, we plot all of the non-zero eigenvalues (left) and the first $200$ non-zero eigenvalues of $\imag{\log{(P e^{2 \pi i X / L_1} P)}}$ (right); notice that the spectral gaps are still exist with the strong disorder. In Figure \ref{fig:hal_per_str_dis_ex}, we plot some of the results of the IPP algorithm.

\FloatBarrier

\subsection{Time Reversal Symmetry Tests}
\label{sec:tr-numerics}
For our last numerical tests, we will test the IPP algorithm for systems with Bosonic time reversal symmetry and Fermionic time reversal symmetry (both $\Z_2$ invariant even and odd cases).

\subsubsection{Haldane Model, Periodic Boundary Conditions, Bosonic Time Reversal Symmetry}
\label{sec:hm-per-bos}
\begin{figure}[h]
  \includegraphics[width=.7\linewidth]{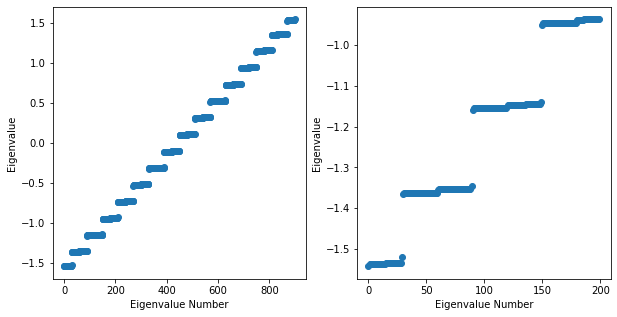}
    \caption{Plot of all non-zero eigenvalues (left) and the first $200$ non-zero eigenvalues (right) of $\arcsin{(P \sin{(2 \pi X / L_1)} P)}$. Here $P$ denotes the Fermi projection for a Haldane model on a $30 \times 30$ system with periodic boundary conditions and $X$ is the standard position operator. The parameters used are $(v, t, t') = (3,1,0.6)$ (see Appendix \ref{sec:model-definitions} for definition of parameters).}
  \label{fig:hal_bos_psinp}
\end{figure}

\begin{figure}[h]
  \includegraphics[width=.75\linewidth]{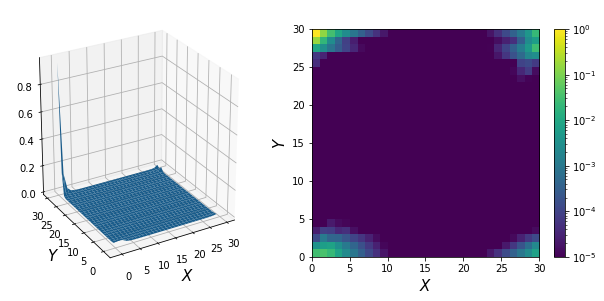} \\
  \includegraphics[width=.75\linewidth]{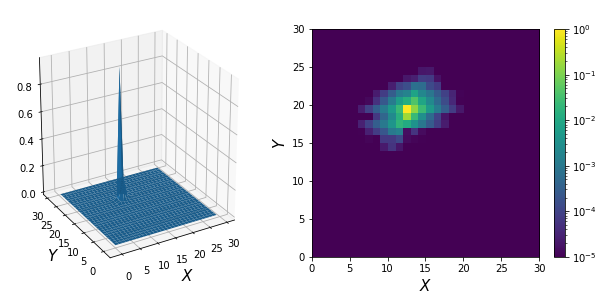}
  \caption{Plot of two of the generalized Wannier functions generated by the IPP algorithm using $\sin{(2 \pi X / L_1))} \rightarrow \cos{(2 \pi X/ L_1))} \rightarrow \sin{(2 \pi Y / L_2)} \rightarrow \cos{(2 \pi Y / L_2)}$ for the system from Figure \ref{fig:km_per_z0_evals_psinp}. A 3D surface plot of the results (left) and the corresponding 2D log plot (right).}
  \label{fig:hal_bos_ex}
\end{figure}

As our first numerical test of the relationship between the IPP algorithm and time reversal symmetry. We consider a Haldane model with parameters $(v, t, t') = (3,1,.5i)$ (where $i = \sqrt{-1}$). In this case, the Haldane Hamiltonian has only real entries and therefore satisfies Bosonic time reversal symmetry. Since the eigenvectors of a Hermitian matrix with real entries can always chosen to be real, it is easy to see that performing the sequence of diagonalizations $\sin{(2 \pi X / L_1))} \rightarrow \cos{(2 \pi X/ L_1))} \rightarrow \sin{(2 \pi Y / L_2)} \rightarrow \cos{(2 \pi Y / L_2)}$ will always generate real Wannier functions without any additional computational effort.  In Figure \ref{fig:hal_bos_psinp}, we plot all of the non-zero eigenvalues (left) and the first $200$ non-zero eigenvalues of $\arcsin{(P \sin{(2 \pi X / L_1)} P)}$ (right). In Figure \ref{fig:hal_bos_ex}, we plot some of the results of the IPP algorithm.

\FloatBarrier
\subsubsection{Kane-Mele Model, Periodic Boundary Conditions, $\Z_2$ invariant even}
\label{sec:km-per-z0}
\begin{figure}[h]
  \includegraphics[width=.7\linewidth]{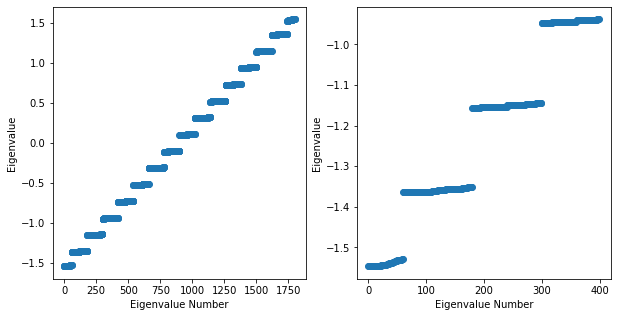}
    \caption{Plot of all non-zero eigenvalues (left) and the first $400$ non-zero eigenvalues (right) of $\arcsin{(P \sin{(2 \pi X / L_1)} P)}$. Here $P$ denotes the Fermi projection for a Kane-Mele model on a $30 \times 30$ system with periodic boundary conditions and $X$ is the standard position operator. The parameters used are $(v, t, t', \lambda_R) = (4,1,0.6,0.5)$ (see Appendix \ref{sec:model-definitions} for definition of parameters).}
  \label{fig:km_per_z0_evals_psinp}
\end{figure}

\begin{figure}[h]
  \includegraphics[width=.75\linewidth]{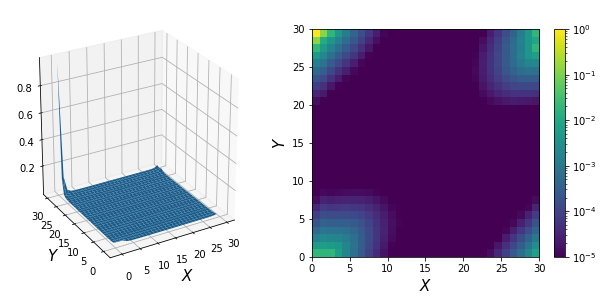} \\
  \includegraphics[width=.75\linewidth]{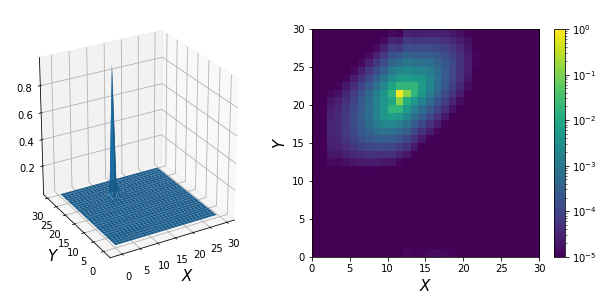}
  \caption{Plot of two of the generalized Wannier functions generated by the IPP algorithm using $\sin{(2 \pi X / L_1))} \rightarrow \cos{(2 \pi X/ L_1))} \rightarrow \sin{(2 \pi Y / L_2)} \rightarrow \cos{(2 \pi Y / L_2)}$ for the system from Figure \ref{fig:km_per_z0_evals_psinp}. A 3D surface plot of the results (left) and the corresponding 2D log plot (right).}
  \label{fig:km_per_z0_ex}
\end{figure}

For our first test of Fermionic time reversal symmetry, let us consider the Kane-Mele model with even $\Z_2$ invariant. For this test we use parameters $(v,t,t',\lambda_R) = (4, 1, 0.6, 0.5)$. In Figure \ref{fig:km_per_z0_evals_psinp}, we plot the non-zero eigenvalues (left) and the first $400$ non-zero eigenvalues of $\arcsin{(P \sin{(2 \pi X / L_1)} P)}$.  In Figure \ref{fig:km_per_z0_ex}, we plot some of the results of the IPP algorithm.

\FloatBarrier
\subsubsection{Kane-Mele Model, Periodic Boundary Conditions, $\Z_2$ invariant odd}
\label{sec:km-per-z1}
For our next test, let us consider the Kane-Mele model with odd $\Z_2$ invariant. For this test we will use parameters $(v,t,t',\lambda_R) = (0, 1, 0.6, 0.5)$.

\begin{figure}[h]
  \includegraphics[width=.7\linewidth]{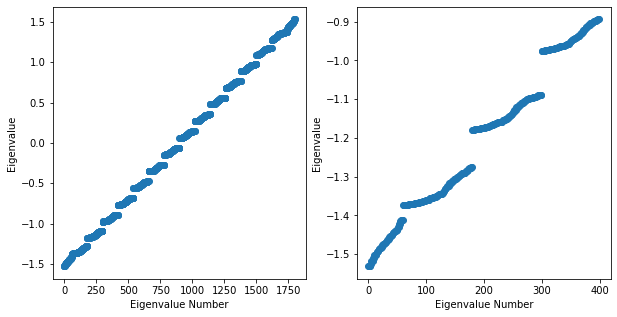}
  \caption{Plot of all non-zero eigenvalues (left) and the first $400$ non-zero eigenvalues (right) of $\arcsin{(P \sin{(2 \pi X_{TRB} / L_1)} P)}$. Here $P$ denotes the Fermi projection for a Kane-Mele model on a $30 \times 30$ system with periodic boundary conditions and $X_{TRB}$ is the ``time reversal broken'' position operator introduced in Section \ref{sec:tr_breaker}. The parameters used are $(v, t, t', \lambda_R) = (0,1,0.6,0.5)$ (see Appendix \ref{sec:model-definitions} for definition of parameters).}
  \label{fig:km_per_z1_evals_psinp}
\end{figure}

As discussed in Section \ref{sec:tr_breaker}, due topological obstructions, the $\Z_2$ invariant odd case requires we modify the choice of position operators to break time reversal symmetry. For these purposes, in Section \ref{sec:tr_breaker} we introduced the time reversal breaker, $A_{TRB}$, and observed that adding $A_{TRB}$ to the standard position operators causes gaps to open in the spectrum of $P \sin{(2 \pi X_{TRB} / L_1)} P$. In Figure \ref{fig:km_per_z1_evals_psinp}, we plot the non-zero eigenvalues (left) and the first $400$ non-zero eigenvalues of $\arcsin{(P \sin{(2 \pi X_{TRB} / L_1)} P)}$. In Figure \ref{fig:km_per_z0_ex}, we plot some of the results of the IPP algorithm.

\begin{figure}[h]
  \includegraphics[width=.75\linewidth]{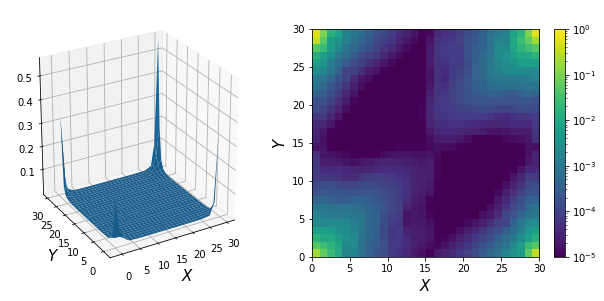} \\
  \includegraphics[width=.75\linewidth]{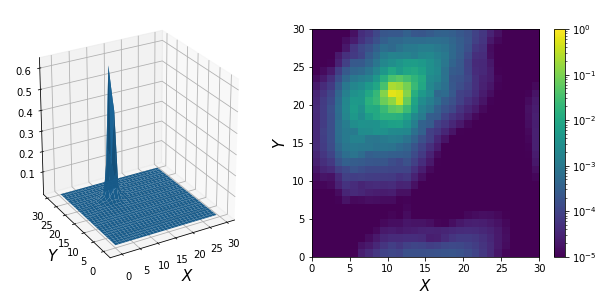}
    \caption{Plot of two of the generalized Wannier functions generated by the IPP algorithm using $\sin{(2 \pi X_{TRB} / L_1))} \rightarrow \cos{(2 \pi X_{TRB} / L_1))} \rightarrow \sin{(2 \pi Y_{TRB} / L_2)} \rightarrow \cos{(2 \pi Y_{TRB} / L_2)}$ for the system from Figure \ref{fig:km_per_z1_evals_psinp}. A 3D surface plot of the results (left) and the corresponding 2D log plot (right).}
  \label{fig:km_per_z1_ex}
\end{figure}

\FloatBarrier
\subsubsection{Kane-Mele Model, Periodic Boundary Conditions, $\Z_2$ invariant odd, Weak Disorder}
\label{sec:km-per-z1-wk-dis}
As noted in Section \ref{sec:hm-per-wk-dis}, any amount of unstructured disorder destroys the periodicity present in the system and therefore Bloch theory cannot be applied. Despite this difficulty, the IPP algorithm remains robust to disorder. For the following tests, the parameters for the Kane-Mele model are $(v,t,t',\lambda_R) = (0, 1,0.6,0.5)$ and the on-site disorder has variance $\sigma^2 = 0.5$.

\begin{figure}[h]
  \includegraphics[width=.7\linewidth]{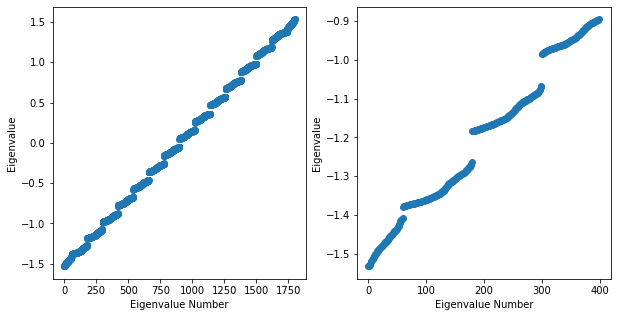}
    \caption{Plot of all non-zero eigenvalues (left) and the first $400$ non-zero eigenvalues (right) of $\arcsin{(P \sin{(2 \pi X_{TRB} / L_1)} P)}$. Here $P$ denotes the Fermi projection for a Kane-Mele model with disorder on a $30 \times 30$ system with periodic boundary conditions and $X_{TRB}$ is the ``time reversal broken'' position operator introduced in Section \ref{sec:tr_breaker}. The parameters used are $(v, t, t', \lambda_R) = (0,1,0.6,0.5)$ and the disorder has variance $\sigma^2 = 0.5$ (see Appendix \ref{sec:model-definitions} for definition of parameters and Equation \eqref{eq:disorder} for definition of on-site disorder).}
  \label{fig:km_per_z1_wk_dis_evals_psinp}
\end{figure}

\begin{figure}[h]
  \includegraphics[width=.75\linewidth]{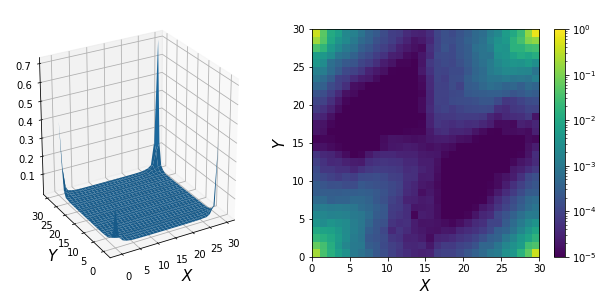} \\
  \includegraphics[width=.75\linewidth]{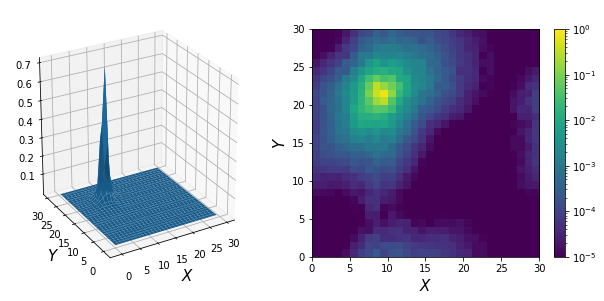}
  \caption{Plot of two of the generalized Wannier functions generated by the IPP algorithm using $\sin{(2 \pi X_{TRB} / L_1))} \rightarrow \cos{(2 \pi X_{TRB} / L_1))} \rightarrow \sin{(2 \pi Y_{TRB} / L_2)} \rightarrow \cos{(2 \pi Y_{TRB} / L_2)}$ for the system from Figure \ref{fig:km_per_z1_wk_dis_evals_psinp}. A 3D surface plot of the results (left) and the corresponding 2D log plot (right).}
\end{figure}

\section{Conclusions}
\label{sec:conclusions}
In this work we have introduced the iterated projected position (IPP) algorithm as an optimization-free method for constructing exponentially localized generalized Wannier functions in both periodic and non-periodic materials in two dimensions and higher. The key assumption underlying the IPP algorithm that $PXP$ has ``uniform spectral gaps'', that the spectrum of $PXP$ can be decomposed into a disjoint union of separated sets. Our previous work \cite{2020StubbsWatsonLu} has shown that if $PXP$ has uniform spectral gaps then an exponentially localized basis for the Fermi projection exists. We have shown that uniform spectral gaps is consistent with previously the known theory on topological invariants. While previous works have considered the projected position operator $PXP$, one key difference for the IPP algorithm is that we can replace the standard position operator $X$ with any local, self-adjoint operator $\tilde{X}$. So long as $P \tilde{X} P$ has uniform spectral gaps, the same theoretical results for $PXP$ also hold for $P \tilde{X} P$. We make use of this freedom in our numerical experiments of the Kane-Mele model with odd $\Z_2$ invariant. For such a model, in agreement with previously known theory, we find that $PXP$ does not have uniform spectral gaps and hence the IPP algorithm fails. To overcome this difficulty, we define a local perturbation $A_{TRB}$ which explictly breaks time reversal symmetry and set $\tilde{X} = X + A_{TRB}$. Once we define $\tilde{X}$ in this way, we find that $P \tilde{X} P$ has uniform spectral gaps and verify that using the IPP algorithm with $P \tilde{X} P$ give functions which are exponentially localized. We conjecture that that this behavior is generically true; that it is \textit{always} possible to construct a modified position operator $\tilde{X}$ so that $P \tilde{X} P$ has uniform spectral gaps so long as a localized basis for the Fermi projection exists. 

\smallskip 
\noindent \textbf{Acknowledgements.}
This work is supported in part by the National Science Foundation via grant DMS-2012286 and the Department of Energy via grant DE-SC0019449. K.D.S. was supported in part by a National Science Foundation Graduate Research Fellowship under Grant No. DGE-1644868. A.B.W. would like to thank Terry A. Loring for helpful discussions.

\bibliography{bibliography}

\FloatBarrier

\appendix
\section{Model Definitions}
\label{sec:model-definitions}
\subsection{Haldane Model}
The Haldane model, first introduced by Haldane in \cite{1988Haldane}, describes electrons in the tight binding model on the honeycomb lattice. The Haldane Hamiltonian with parameters $(v, t, t')$ can be written as follows:
\begin{equation}
  \label{eq:hal-h-def}
  H_{Hal}  = v \sum_{j} \xi_j c_j^\dagger c_j +  t \sum_{<jk>} c_j^\dagger c_k   + i t' \sum_{\ll jk \gg} \nu_{jk}  c_j^\dagger c_k
\end{equation}
In Equation \ref{eq:hal-h-def}
\begin{itemize}[itemsep=-.5ex,label=--]
\item $c_j$ is the annihilation operator at site $j$.
\item $v$ is the onsite potential difference.
\item $\xi_j$ takes the value $+1$ on $A$ sites and $-1$ on $B$ sites.
\item $t$ is the nearest neighbor hopping amplitude.
\item $t'$ is the next nearest neighbor hopping amplitude.
\item $\nu_{jk}$ is $\pm 1$ depending on the relative orientation between sites $j$ and $k$.
\end{itemize}
  
\subsection{Kane-Mele Model}
The Kane-Mele model, first introduced by Kane and Mele in \cite{2005KaneMele}, generalizes the Haldane model to include spin with time reversal invariant spin orbit interations. The Kane-Mele Hamiltonian with parameters $(v, t, t', \lambda_R)$ can be written as follows:
\begin{equation}
  \label{eq:km-h-def}
  \begin{split}
    H_{KM} = & v \sum_{j} \xi_j \vec{c}_{j}^\dagger \vec{c}_{j} +  t \sum_{<jk>} \vec{c}_{j}^\dagger \vec{c}_{k} \\
    & + i t' \sum_{\ll jk \gg} \nu_{jk}  \vec{c}_{j}^\dagger \sigma_z \vec{c}_{k} + i \lambda_R \sum_{<jk>} \vec{c}_{j}^\dagger ( \vec{s} \times \vec{d}_{jk})_z \vec{c}_{k}
  \end{split}
\end{equation}
In Equation \ref{eq:km-h-def}
\begin{itemize}[itemsep=-.5ex,label=--]
\item $\vec{c}_j$ is the Fermionic annihilation operator at site $j$.
\item $v$ is the onsite potential difference.
\item $\xi_j$ takes the value $+1$ on $A$ sites and $-1$ on $B$ sites.
\item $t$ is the nearest neighbor hopping amplitude.
\item $t'$ is the next nearest neighbor hopping amplitude.
\item $\nu_{jk}$ is $\pm 1$ depending on the relative orientation between sites $j$ and $k$.
\item $\lambda_R$ is the strength of the Rashba interaction.
\item $\vec{s}$ is a vector of Pauli matrices.
\item $\vec{d}_{jk}$ is vector pointing from site $j$ to $k$.
\end{itemize}

\subsection{$p_x + i p_y$ Model}
The $p_x + i p_y$ model, first introduced by Fulga, Pikulin, and Loring in \cite{2016FulgaPikulinLoring}, was developed as an example for an aperiodic system which can host weak topological phase. The $p_x + i p_y$ Hamiltonian with parameters $(\mu, t, \Delta)$ can be written in terms of the $2 \times 2$ matrices $H_i$ and $H_{jk}$.
\[
  \begin{array}{l}
    \displaystyle H_j = - \mu \sigma_z \\[1ex]
    \displaystyle H_{jk}  = -t \sigma_z - \frac{i\Delta}{2} \cos{(\alpha_{jk})} \sigma_x - \frac{i\Delta}{2} \sin{(\alpha_{jk})} \sigma_y \\
  \end{array}
\]
Given these definitions the $p_x + ip_y$ Hamiltonian can be written as:

\begin{equation}
  \label{eq:pxipy-h-def}
  H_{p_x+ip_y} = \sum_{j} \vec{c}_j^\dagger H_j \vec{c}_j + \sum_{<jk>} \vec{c}_{j}^\dagger H_{jk} \vec{c}_{k}
\end{equation}
In Equation \eqref{eq:pxipy-h-def}
\begin{itemize}[itemsep=-.5ex,label=--]
\item $\vec{c}_j$ is the Fermionic annihilation operator at site $j$.
\item $\mu$ is the chemical potential.
\item $t$ is the hopping strength between neighboring sites.
\item $\Delta$ is the strength of the $p$-wave pairing.
\item $\alpha_{jk}$ is the angle of the  bond  between site $j$ and site $k$ measured with respect to the horizontal direction.
\end{itemize}



\begin{figure}
  \centering
  \includegraphics[width=.6\linewidth]{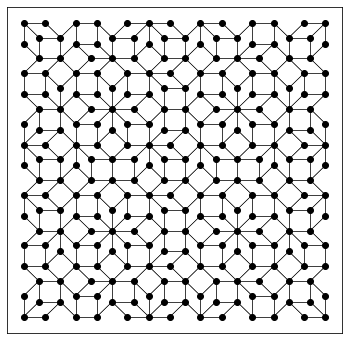}
  \caption{An Ammann-Beekner tiling of a square domain.}
  \label{fig:ammann-beekner}
\end{figure}

\section{Calculation for the Marzari-Vanderbilt Functional}
\label{sec:mv-functional-calc}
In this section we will show the equivalence between Equation \eqref{eq:mv-functional-modified} and the gauge dependent part of the Marzari-Vanderbilt functional from \cite{1997MarzariVanderbilt}. These calculations essentially rederive Equations (15) and (16) in \cite{1997MarzariVanderbilt} using different notation.

For this calculation, recall $P$ denotes the Fermi projection and let us define $Q = I - P$. Since $P$ is a projection we have that $PQ = QP = 0$. Also, recall that we define
\[
  \mu_{n\vec{0}}^X := \bra{w_{n\vec{0}}} X \ket{w_{n\vec{0}}} \qquad \mu_{n\vec{0}}^Y := \bra{w_{n\vec{0}}} X \ket{w_{n\vec{0}}}.
\]

By definition the variance in the $X$ direction of $w_{n\vec{0}}$ can be written as:
\begin{align*}
  \text{Var}_X(w_{n\vec{0}}) & = \int (x - \mu_{n\vec{0}}^X)^2 |w_{n\vec{0}}(x,y)|^2 \dee{x} \dee{y} \\
  & = \bra{w_{n\vec{0}}} (X - \mu_{n\vec{0}}^X)^2 \ket{w_{n\vec{0}}} \\
  & = \bra{w_{n\vec{0}}}P (X - \mu_{n\vec{0}}^X)^2 P \ket{w_{n\vec{0}}}
\end{align*}
Using the fact that $P + Q = I$ and $PQ = QP = 0$, we can rewrite the operator $P (X - \mu)^2 P$ as follows:
\begin{align*}
  P(X &- \mu)^2P  = P(X - \mu)(P+Q)(X - \mu)P \\
      & = P(X - \mu)P(X - \mu)P + P(X - \mu)Q(X-\mu)P \\
      & = P(X - \mu)PP(X - \mu)P + PXQXP \\
      & = (P(X - \mu)P)^2 + (QXP)^\dagger QXP 
\end{align*}
Therefore, using that $\bra{w} A^\dagger A \ket{w} = \| A w\|^2$ we have that
\[
  \text{Var}_X(w_{n\vec{0}}) = \| P(X - \mu_{n\vec{0}}^X)P w_{n\vec{0}} \|^2 + \| QXP w_{n\vec{0}} \|^2.
\]
A similar calculation in $Y$ shows that
\[
  \text{Var}_Y(w_{n\vec{0}}) = \| P(Y - \mu_{n\vec{0}}^Y)P w_{n\vec{0}} \|^2 + \| QYP w_{n\vec{0}} \|^2.
\]
Now taking the sum of the variance over the bands gives us that the Marzari-Vanderbilt functional is:
\begin{align*}
  F_{MV}(w) & = \sum_{n} \text{Var}_X(w_{n\vec{0}}) + \text{Var}_Y(w_{n\vec{0}}) \\
         & = \sum_{n} \| P(X - \mu_{n\vec{0}}^X)P w_{n\vec{0}} \|^2  + \| P(Y - \mu_{n\vec{0}}^Y)P w_{n\vec{0}} \|^2 \\
  & ~~~~~~+ \sum_{n} \| QXP w_{n\vec{0}} \|^2 + \| QYP w_{n\vec{0}} \|^2
\end{align*}
Since $\{ w_{n\vec{0}} \}$ forms a basis for the Fermi projection over the unit cell, the term
\[
  \sum_{n} \| QXP w_{n\vec{0}} \|^2 + \| QYP w_{n\vec{0}} \|^2
\]
is just the sum of the traces over the unit cell of the operators $PXQXP$ and $PYQYP$. Since the trace is independent of basis, these two terms are independent of gauge (cf. Equation (16) in \cite{1997MarzariVanderbilt}).

To see that the remaining terms correspond to the gauge dependent part of the Marzari-Vanderbilt functional (Equation (15) in \cite{1997MarzariVanderbilt}) recall that the Wannier functions $\{ w_{m\vec{R}} \}$ form an orthogonal basis for $\range{(P)}$. Using this fact we have that:
\begin{align*}
  \| P(X -& \mu_{n\vec{0}}^X) P w_{n\vec{0}} \|^2 = \sum_{m\vec{R}} |\bra{w_{m\vec{R}}} (X - \mu_{n\vec{0}}^X)P \ket{w_{n\vec{0}}}|^2 \\
          &  = \sum_{m\vec{R}} \left|\bra{w_{m\vec{R}}} X \ket{w_{n\vec{0}}} - \mu_{n\vec{0}}^X \bra{w_{m\vec{R}}} \ketl{w_{0\vec{n}}}\right|^2 \\
          &  = \sum_{m\vec{R}\neq n\vec{0}} \left|\bra{w_{m\vec{R}}} X \ket{w_{n\vec{0}}}\right|^2
\end{align*}
where we have used that $\bra{w_{m\vec{R}}} \ketl{w_{0\vec{n}}} = \delta_{m\vec{R},0\vec{n}}$.

\end{document}